\documentclass[journal]{IEEEtran}
\usepackage{cite}
\usepackage{graphicx}
\usepackage{subfigure}
\usepackage{amsmath}
\usepackage{amsfonts,amssymb}
\usepackage{algorithm}
\usepackage{algorithmic}
\usepackage{array}
\usepackage{multirow}
\makeatletter
\newcommand{\removelatexerror}{\let\@latex@error\@gobble}
\makeatother

\begin{document}
\title{Automatic Radio Map Adaptation for Robust Localization with Dynamic Adversarial Learning}
% \title{Cross-Scenario Device-Free Localization Based on Dynamic Adversarial Learning}

\author{
Lingyan~Zhang,~Junlin~Huang,~Tingting~Zhang,~\IEEEmembership{Member,~IEEE,} and  Qinyu~Zhang,~\IEEEmembership{Senior Member,~IEEE}
        % Michael~Shell,~\IEEEmembership{Member,~IEEE,}

% \thanks{This work has been supported in part by the National Natural Science Foundation of China under Grant No. ~62171160 and ~61771159, in part by China Postdoctoral Science Foundation under Grant No.~2020M670910, and part by Shenzhen Science and Technology Program under Grant No.~JCYJ20190806143212658 and ZDSYS20210623091808025. }
%\thanks{Lingyan Zhang is with the Department of Electronic and Electrical Engineering, Southern University of Science and Technology, Shenzhen 518055, China  (e-mail: lingyanz2020@gmail.com)}
%\thanks{Shaohua. Wu, Tingting. Zhang, and Qinyu. Zhang are with the School of Electronics and Information Engineering, Harbin Institute of Technology (Shenzhen), Shenzhen 518055, China (e-mail:  hitwush@hit.edu.cn,  zhangtt@hit.edu.cn, and zqy@hit.edu.cn). }
% <-this % stops a space
%\thanks{}% <-this % stops a space
%\thanks{}% <-this % stops a space
% \thanks{}% <-this % stops a space
}

%\markboth{IEEE Internet of Things Journal}
% \markboth{	IEEE Transactions on Industrial Informatics}
% {Shell \MakeLowercase{\textit{et al.}}: }

\maketitle

\begin{abstract}
Wireless fingerprint-based localization has become one of the most promising technologies for ubiquitous location-aware computing and intelligent location-based services. However, due to RF vulnerability to environmental dynamics over time, continuous radio map updates are time-consuming and infeasible, resulting in severe accuracy degradation. To address this issue, we propose a novel approach of robust localization with dynamic adversarial learning, known as DadLoc which realizes automatic radio map adaptation by incorporating multiple robust factors underlying RF fingerprints to learn the evolving feature representation with the complicated environmental dynamics. DadLoc performs a finer-grained distribution adaptation with the developed dynamic adversarial adaptation network and quantifies the contributions of both global and local distribution adaptation in a dynamics-adaptive manner. Furthermore,  we adopt the strategy of prediction uncertainty suppression to conduct source-supervised training, target-unsupervised training, and source-target dynamic adversarial adaptation which can trade off the environment adaptability and the location discriminability of the learned deep representation for safe and effective feature transfer across different environments.  With extensive experimental results, the satisfactory accuracy over other comparative schemes demonstrates that the proposed DanLoc can facilitate fingerprint-based localization for wide deployments.
\end{abstract}


\IEEEpeerreviewmaketitle

\section{Introduction}
Recently,  wireless fingerprint-based localization has become one of the most appealing technologies to provide location-aware computing and location-based services (LBS) for various practical and potential applications \cite{IPSSurvey,AIWiFi}, such as 
accurate localization and navigation, smart home, geo-based augmented/virtual reality, mobile ads targeting, autonomous  driving, industrial internet, and so on. 
%It usually refers to machine learning to estimate users' locations by optimally fingerprinting with  a radio map that associates the signal features with the corresponding location information.
With the ubiquitous penetration of WLAN in our surrounding environments,
% various wireless techniques, such as WiFi \cite{WiFi, DeFi,3DWiFi}, UWB \cite{STUWB,SUWB}, RFID \cite{RfidDude,RF-CHORD}, visible lights \cite{vlight}, and millimeter waves \cite{mmtrack}, have been fully exploited to design and implement indoor localization systems for providing LBS.   
WiFi fingerprint-based localization \cite{FLSurvey,  AIWiFi,AuMc} has
many advantages over others \cite{SUWB,3DWiFi,DeFi}, as results of easy signal acquisition, simple localization algorithm, and satisfactory system performance, especially deep-learning-based schemes \cite{DLoc,l2l,WiDeep,CNNLoc,DeepFi,RNN} which learn implicitly deep features to  characterize the intrinsic mappings between signal features and ground-truth locations, effectively attaining accuracy improvement. 
However, classical deep-neural-network-based localization schemes \cite{CNNLoc,WiDeep,DeepFi} have extracted  signal features to construct a static radio map without the adaptability of environmental dynamics. 
Practically,  radio signals are extremely vulnerable to environmental changes which can lead to
feature shifts and further cause the mismatch between the existing static radio map and current signal features, resulting in severe accuracy degradation, even a failure of localization \cite{AuMc,FLSurvey,l2l}.

For robust  location estimations, the radio map  is intuitively required for continuous calibration or reconstruction with up-to-date RF fingerprints \cite{AuMc,TRAN,LSTMLoc}. 
%Instead of assisted devices and auxiliary measurements with crowdsourcing sensing\cite{AuMc,Fincs}, 
Advance works propose the algorithm-level schemes on radio map adaptation at low maintenance in a framework of transfer learning which attempts to reduce feature  shifts before and after the environmental changes \cite{TCA,AIWiFi,FLSurvey}.
The kernel-based methods \cite{TCA,TKL,MTL,TRAN,LSTMLoc} are adopted
to minimize the measurable distances of feature distribution in the latent space, and then 
the common shared features are extracted as transferable knowledge which is acquired at
high and complex computational cost for users' location estimations in new environments. 
%, while they are acquired at high and complex computational cost. 
%Moreover,  the environmental dynamics  are mixed with both the long-term weather changes and the short-term fluctuation of object or target motion \cite{LSTMLoc}, however,  these schemes only perform the  distribution alignment of global features   from all reference points cross environments, leading to coarse location estimations suffering from  ineffective transfer  \cite{ARTL,MEDA}. 
Alternatively,  pioneer schemes \cite{ILOT,DRF}  incorporate deep neural network (DNN) with domain adversarial learning 
to learn environment-invariant representations by maximizing the confusion of feature distributions across different environments (domains), while the domain discriminator is trained to minimize the classification error of differentiating the source fingerprint from the target data. 
%By developing adversarial adaptation networks \cite{DAAN,DA}, the domain discriminator is trained to minimize the classification error of differentiating the source from the target environments,  meanwhile the  location prediction model learns transferable representation to deceive the domain discriminator. 
Instead of other assisted devices and auxiliary measurements, such as crowdsourcing sensing\cite{AuMc,Fincs}, these works can retain the ubiquity to facilitate robust fingerprint-based localization to provide a wide spread of LBS. 

Although adversarial-adaptation-based schemes effectively strengthen the transferability of feature representations which can bridge the discrepancy across environments, 
%its impact on the feature discriminability, as a key factor of the location determination, has not fully explored and exploited to achieve coarse-grained accuracy.
% on account of  overlooking the inherent location information of RF fingerprints. 
%Therefore, the evolving signal representation should be learned to  boost  the feature discriminability  for gaining accuracy improvement. 
there are the following limitations to enable current robust localization as fully practical LBS. Firstly,  many works  make sparse spatial constraints to only perform the feature distribution adaptation without the inherent location information of RF fingerprints which leads to  coarse location estimations suffering from \textbf{ineffective transfer}. 
Due to  multipath fading, temperature or humidity changes, and transient interference with the object or user movement in the interested environments, 
complex and multivariate variances are involved in the effectiveness of transferring feature representation. 
A single robust factor underlying both source  fingerprints and target RF measurements is  insufficient enough to learn dynamics-resistant feature representation for accurate  localization  with complicated environmental dynamics \cite{ARTL,MEDA}. 
Secondly, current solutions cannot fully react to real-time environmental dynamics, and thus the \textbf{environmental adaptability}  should be enhanced to accommodate diverse environmental dynamics at any time. On providing LBS, the continuous changes and unpredictable interference in the interested environment cause 
 the cumulative variances of radio signals over time to large localization errors \cite{LSTMLoc,AuMc}. 
% \textbf{environmental adaptability}  should be enhanced to accommodate diverse environmental dynamics at any time. The learned adaptation model should enable dynamic radio map adaptation to accommodate diverse environmental changes.  
%It requires on-time radio map adaptation to for offering a long-term LBS at a pretty low maintenance cost.
Finally, 
the \textbf{location discriminability} of transferable feature representations, as a key of the location determination, has not been fully  exploited to yield reasonable performance gains. During adversarial learning,  transferable knowledge across environments is acquired by the maximization of  feature distribution confusion which weakens their location discriminability  to achieve unsatisfactory location estimations with complicated environmental dynamics. 

To address these issues comprehensively, we propose a novel approach of robust fingerprint-based localization with 
 \textbf{d}ynamic  \textbf{ad}versarial adaptation, as DadLoc which 
enables automatic radio map adaptation to achieve accurate location estimations with the robustness of complicated environmental dynamics.  
DadLoc incorporates multiple robust factors underlying RF fingerprints from different environments to learn the evolving feature representation with dynamic adversarial learning, which  trades off environmental adaptability and location discriminability.  
%A finer-grained distribution adaptation is performed to learn transferable feature representation by incorporating multiple factors of effective knowledge transfer. 
Specifically, 
we develop dynamic adversarial adaptation network (DAAN) to remove both feature (global) and fingerprint (local) distribution discrepancies across different environments in a  dynamics-adaptive manner which automatically weighs the importance of global and local distribution adaptation with the environmental changes at any time. % We fully use the unlabeled RF samples to update DAAN, and further the evolving feature representations are learned to effectively acquire fine environmental adaptability. 
Furthermore, we investigate the key cause of coarse location estimations  and map out the training strategy of prediction uncertainty suppression to gain the effectiveness of accuracy improvement. 
According to the path loss model\cite{LPLM}, the characteristics of RF propagation are disclosed that there are different capabilities of signal representation from different locations.  
%An identical RF fading incurs a smaller distance variance at closer locations to the transmitter, whereas,
Far from WiFi AP, RF fingerprints   have weaker  representation capability with higher location uncertainty to  result in larger location prediction errors, which can be exacerbated with complicated environmental dynamics. Under domain adversarial adaptation, it is highly risky to learn transferable feature representation by minimizing the distribution discrepancies, even if the feature distribution confusion is maximized \cite{DDAN, CDAN}. 
Toward effective and safe transfer, 
DadLoc quantitatively evaluates the location uncertainty of RF fingerprints and removes this negative impact on dynamic distribution adaptation to learn finer transferable representation for yielding remarkable performance gains.
% Therefore,
% DadLoc quantitatively evaluates the location uncertainty of original RF fingerprints and adopts 
%  the uncertainty-aware weighting  to emphasize the strong location discriminability  and encourage target location prediction  with high confidence. 
Finally, DadLoc synthesizes the positive transfer factors to train the developed location prediction model with source-supervised training, target-unsupervised training, and source-target dynamic adversarial adaptation.
%Due to the complexity and uncertainty of wireless propagation in indoor scenarios, 
%it is difficult to 
%balance their contributions to accommodate diverse indoor environmental changes. DadLoc can  by   which indicate their respective contributions to realize automatic radio map adaptation. 
% Furthermore, the domain discriminator is further conditioned by quantifying  the uncertainty of location label prediction toward effective knowledge transfer.  
%Furthermore, due to the uncertainty of the location prediction with local subdomain adaptation, we adopt an entropy-aware metric to condition the domain discriminator for effective and safe transfer. 

We prototype the proposed DadLoc system and carry out real-world experiments in many typical indoor scenarios for a long period of more than 8 months. With the extensive experimental results,  DadLoc can achieve better localization performance than others, which facilitates fingerprint-based localization for providing fully practical LBS. 
In summary, the main contributions to the proposed DadLoc system are
represented as follows.
\begin{enumerate}
	\item We propose a novel approach of self-adaptive localization with dynamic adversarial learning to acquire finer transferable feature representation which can further effectively enable robust location estimations at any time with diverse environmental dynamics.  
	
	\item We incorporate multi-level factors of effective  transfer to learn the evolving knowledge by the tradeoff between environmental adaptability and location discriminability during dynamic adversarial adaptation. The proposed DadLoc scheme can narrow the gap between automatic radio map adaptation and localization performance at a low cost. 
 %realize automatic radio map adaptation to dynamically accommodate indoor environmental changes with accurate location estimations, which benefit from the developed dynamic domain adversarial networks with the importance evaluation of both global and local domain adaptation and the uncertain evaluation of the location prediction.  

 % at low maintenance cost in a dynamic self-adaptive manner, which can enhance the effectiveness of the environment-invariant knowledge transfer for location accuracy improvement.
 %and further facilitates  the knowledge-transfer-assisted robust localization for wide deployment.

	\item We validate the effectiveness of the proposed DadLoc system in real  environments with extensive experiments.  Satisfactory results demonstrate that DadLoc can facilitate deep-learning-based localization as fully practical LBS with the elegant ubiquity of RF fingerprinting. 
\end{enumerate}

The rest of this paper is structured as follows. Section II summarizes related work. Section III introduces the problem formulation and our key idea. 
The system design and implementation of the proposed DadLoc system are detailed in 
Section  IV. We then demonstrate the effectiveness of the DadLoc system with an extensive experimental study in Section V. Finally, we conclude our work in Section VI.

\section{Related Work}
We briefly categorize deep-learning-based localization into \textit{radio map construction} by training deep features, and \textit{radio map adaptation} with the robustness of environmental dynamics. 

\subsection{Radio Map Construction}

WiFi fingerprint-based localization has gained great accuracy improvement in a deep-learning-based framework \cite{DeepFi,WiDeep,CNNLoc,DLoc} which can learn the implicit signal representation of RF propagation with  unpredictable reflection, diffraction, scattering, and shadowing in an indoor multipath environment, for radio map construction. In an unsupervised learning manner, DeepFi \cite{DeepFi} uses all optimal weights of deep neural networks (DNN) as fingerprints, then the location estimation is determined in a probabilistic method. Recurrent neural networks (RNN) \cite{RNN} are developed to learn temporal and spatial signal representations for  motion tracking with sequential location estimations. WiDeep \cite{WiDeep} adopts  stacked denoising auto-encoders to extract noise-tolerant signal features and to learn the non-linear correlated relation between different RSS at different locations. CNNLoc \cite{CNNLoc}
incorporates stacked autoencoder with one-dimensional conventional neural networks (CNN)  to learn spatial-related deep representation for accurate location estimations. Furthermore, 
by transforming  multi-antenna and multi-subcarrier CSI measurements into 2D radio images, DLoc \cite{DLoc} fully utilizes  the strong representation capability of  CNN to effectively facilitate the accuracy of fingerprint-based localization.   

These deep-learning-based localization approaches, however, have trained appropriate deep features to construct the static radio map, leading to the limitation of environmental dynamics. In practice,  radio signals are extremely vulnerable to environmental changes. Such RF signal variances incur feature shifts and fingerprint shifts to make an initial radio map  invalid, as a result of inaccurate location estimations, even a failure of localization in a new environment.

\subsection{Radio Map Adaptation}

%Fingerprint-based localization has always attracted extensive interest with many superiorities over other. DadLoc is proposed to overcome the vulnerability of environmental dynamics for robust localization, which involves the following related research.
%Early  works \cite{Sensor, LEASE} deployed a large number of reference anchors to calibrate online RSS measurements for  radio map updates, which are expensive for system scalability to provide  a wide range of LBS applications. 
Recent fingerprint-based localization schemes propose algorithmic adaptation at the low cost of deployment and maintenance, especially in a  deep-learning-based framework. 
%Observation based on practical RF measurements has revealed that although RSS fluctuations at each reference anchor are unpredictable with various environmental changes, the relationship of RSS variances among neighboring fingerprints can remain relatively stable, and thus  AcMu \cite{AuMc} and LEMT \cite{LEMT} establish the functional relationship model  for automatic radio map adaptation. Moreover, the deep-learning-based models are designed to learn transferable representations underlying RF fingerprints. 
%Graph convolutional network  \cite{GCN} is employed to aggregate CSI features, and then the graph attention mechanism is adopted to acquire robust channel  knowledge. 
The Siamese network \cite{Siamese} is designed to extract deep CSI features and  compare their similarities which can characterize the  spatial correlations of CSI data at different RPs.  The deep fuzzy forest is leveraged to train the deep tree model for accurate location predictions \cite{fuzzy}.
LESS \cite{l2l} leverages few-shot relation learning to  model the relationship representations  among the neighborhood fingerprints and achieve satisfactory localization performances. We find out that 
it is crucial  to discover the invariant or stable property underlying RF fingerprints across different environments for radio map adaptation.  %radio maps with environmental dynamics, based on ``maintaining the status quo". 

%Therefore, advance works prefer domain adaptation to achieve radio map adaptation  \cite{TCA,TKL,MTL} by  reducing the distribution shifts between source and target environments. 
%Metric-learning-based localization scheme \cite{MTL} relies on  the Hilbert-Schmidt independence criterion to measure the similarity of different domains, and  then the proper  metrics from the source domain are selected for target  location estimation by the  minimization of the distribution discrepancy. 
%TKL-WinSMS \cite{TKL}  minimizes RSS distribution discrepancy between two different radio maps and identifies the shared characteristics with transfer kernel learning.  These  systems  perform global domain adaptation by alleviating the marginal distribution  difference for user location estimations in new environments. However, without taking the location information into account, they are prone to the underfitting of  target data with ineffective transfer \cite{DA,DAAN}. Therefore, it is indispensable for accurate location prediction to further perform  finer-grained domain adaptation with more robust factors underlying RF fingerprints. 
%with the conditional distribution alignment between the source and target environments. 
 
% In this case, \textit{RobLoc} simultaneously performs geometrical and statistical alignment for dynamic domain adaptation to learn  finer transferable knowledge with the adaptability and generalizability  improvement of the fingerprint-based localization system.  

The state-of-the-art works develop deep adaptation networks to learn transferable feature representation  with the robustness of environmental dynamics. Fidora \cite{Fidora} designs a domain-adaptive classifier to self-calibrate the feature extractor with joint optimization of classification and reconstruction. 
Additionally, advanced
 fingerprint-based localization schemes \cite{DRF,ILOT} are proposed to learn environment-invariant feature representations by incorporating domain adversarial learning and DNN in a two-layer minimax game.
TransLoc \cite{DRF} harnesses deep neural forest to learn deep representations with the feature extractor which is trained and fixed by the source database,  and then   adversarial learning is performed to fine-tune   deep adaptation network by minimizing the representation distance between both the source and target database. These schemes endeavor to strengthen the adaptability of feature representation with environmental dynamics, however, the feature discriminability has not been fully explored and exploited to lead to coarse location estimations.  

On these bases, DadLoc can mind the gap between automatic radio map adaptation and localization accuracy with dynamic  adversarial learning which can perform a finer-grained adversarial adaptation  in a  dynamics-adaptive manner. 
The evolving feature representations are learned to achieve robust localization with the tradeoff between
 the environmental adaptability and the location discriminability of feature representation with complicated environmental changes.

\section{Problem Formulation and Motivation}

%Deep-domain-adaptation-based localization schemes \cite{DRF, ILOT} formally follow the framework of a 

The fingerprint-based localization approach adopts machine learning to estimate user locations which formally collects  extensive RF fingerprints to construct a radio map in an offline training stage. The location estimation with RF queries is determined with best-fitted fingerprinting on an online localization stage. 
 The initial radio map, as source fingerprint database (domain) $\mathcal{S}=\left\{\left(\mathbf{x}_{i}^{s}, y_{i}^{s}\right)\right\}_{i=1}^{n_{s}}$,   is composed of RF signals $\mathbf{x}_i$ with the corresponding location label $y_i$. In a new environment,  the target  data $\mathcal{T}=\left\{\mathbf{x}_{i}^{t}\right\}_{i=1}^{n_t}$ are recorded without the location information. 
With the environmental changes, the feature shifts $P_s(\mathbf{x}^s)\neq P_t(\mathbf{x}^t)$ are mainly caused by humidity, temperature, light,  weather conditions, and so on for the long-term LBS.  
 %there are the gradual changing caused with humidity, temperature, light, and weather conditions for a long time, to the feature shifts which can be formulated as global domain discrepancy $P_s(\mathbf{x}^s)\neq P_t(\mathbf{x}^t)$.
 Meanwhile, RF fluctuations are strong with short-term unpredictable changes, such as user motion, a door opening or closing,  or furniture movement, which incur fingerprint shifts $Q_s(\mathbf{x}^s|y)\neq Q_t(\mathbf{x}^t|y)$  at some specific RPs. With multipath propagation in an interested environment, RF variances are severely exacerbated with complex multivariate factors, resulting in inaccurate location estimations. 
 In order to overcome RF vulnerability of complicated environmental dynamics, our goal  is to learn a deep adaptation model on location prediction with effectively  transferable feature representation for  robust localization.

Currently, adversarial-adaptation-based schemes  \cite{ILOT,DRF} attempt to remove the discrepancy of feature distributions  across domains to learn environment-invariant feature representation for robust location prediction.
After feature distribution matching, \textit{i.e.}, global distribution adaptation illustrated in Fig. 1(b),  
the confusion of feature distributions is conducted, but we can
not differentiate the specific features from which RPs, weakening the discriminative information of feature representation. 
It is quite necessary to propose a finer-grained distribution adaptation across different environments to remove the distribution discrepancy for robust location estimation.

Practically, the environmental dynamics at any time can cause different influences on feature distribution and fingerprint distribution. If the deep adaptation model can accommodate the environmental changes in a dynamic self-adaptive manner, the effectiveness enhancement of learned transferable representation can gain localization performance improvement. 
As illustrated by Fig. 1, dynamic distribution adaptation should be finally carried out to learn the evolving feature representation for accurate location prediction with the robustness of environmental changes. 

\begin{figure}[tbp]
	\centering
 \includegraphics[width=0.49\textwidth]{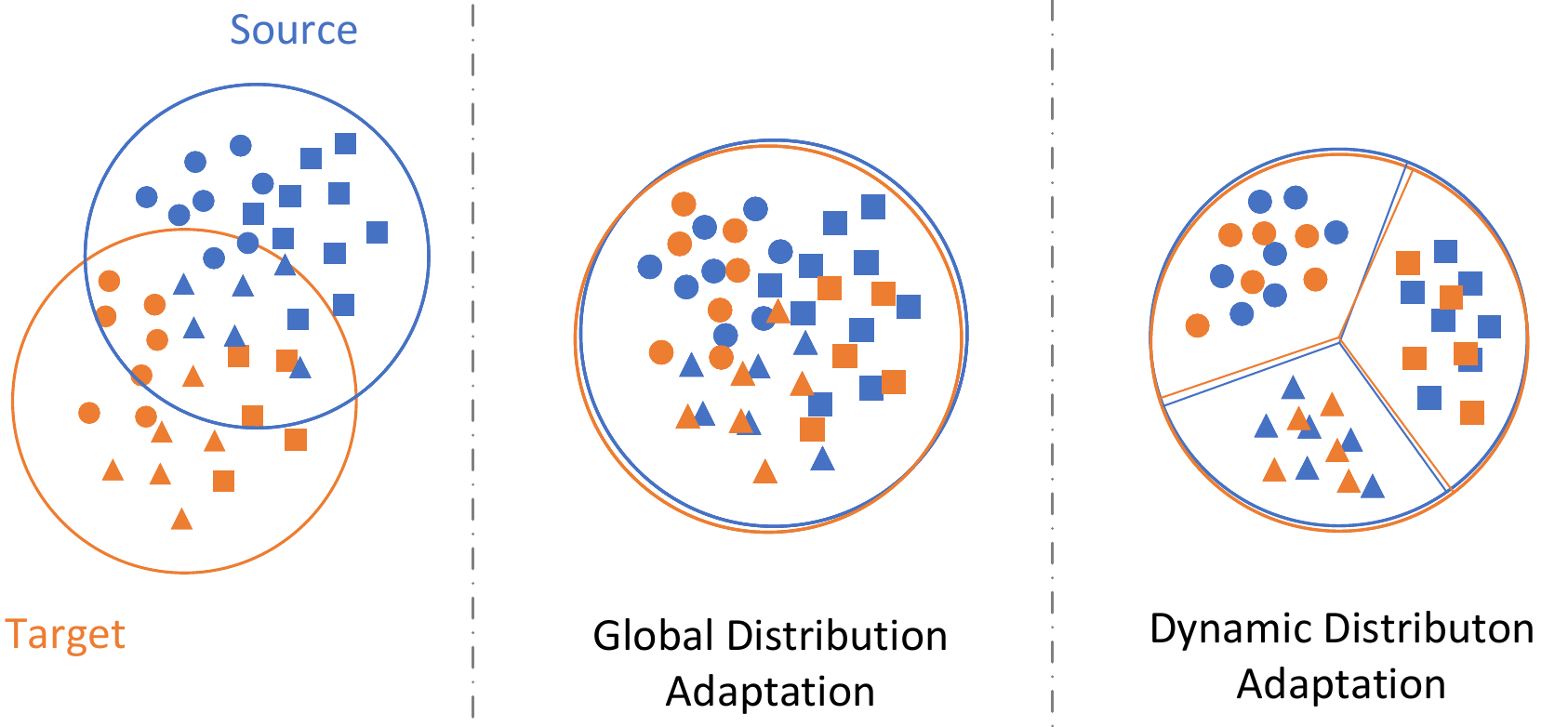}
	\caption{Comparison of global distribution adaptation and dynamic distribution adaptation. }
	%\label{fig_sa}
\end{figure}

%Signal attenuation, multipath effect, shadow fading

% \begin{equation}
%     \mathbf{A}=\left[\begin{array}{cc}
% x_{1} & y_{1} \\
% x_{2} & y_{2} \\
% \vdots & \vdots \\
% x_{T} & y_{T} 
% \end{array}\right]_{T \times 2}
% \end{equation}

%\section{ Self-Adaptive Localization with Manifold Embedded Dynamic Adaptation Networks}

\section{Dynamic   Adversarial Adaptation Based Robust Localization } 

In this section, we elaborate on the system design and implementation of the proposed DadLoc scheme. Firstly, we present DadLoc's system overview and investigate the root cause of coarse localization. Furthermore, we introduce the details of the dynamic adversarial adaptation network to learn finer transferable representation for automatic radio map adaptation. Simultaneously, the strategy of prediction uncertainty suppression is integrated to balance the environmental adaptability and  the location discriminability of feature representations through source-supervised training, target-unsupervised training, and source-target dynamic adversarial learning. 

\subsection{System Overview}

\begin{figure}[tbp]
\centering
\includegraphics[width=0.48\textwidth]{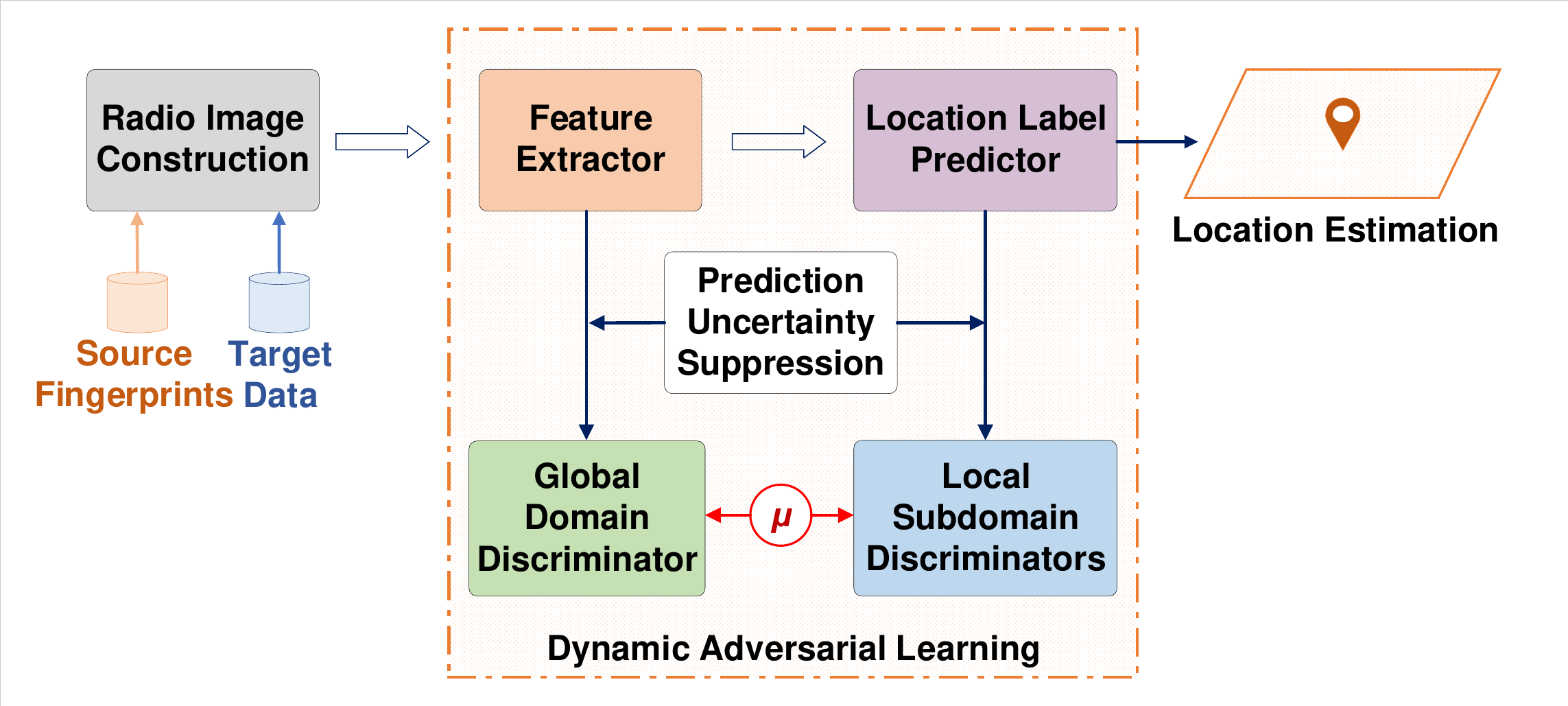}
\caption{System Overview}
\label{fig_sa}
\end{figure}

As an illustration of the system overview in Fig. \ref{fig_sa}, DadLoc follows the deep learning framework of fingerprint-based localization, and then incorporates dynamic adversarial learning to perform finer-grained distribution adaptation for robust localization. 

DadLoc collects WiFi 
CSI fingerprints to construct the source database. We transform the CSI values into radio images, inspired by the strong representation capability of CNN \cite{DLoc,ILOT}. The location prediction model  including the \textit{feature extractor} and the \textit{location  predictor} is initially trained with these source CSI fingerprints.
To acquire environmental adaptability, we perform a finer-grained distribution adaptation in a dynamics-adaptive manner to learn effectively transferable representation by enabling both feature and fingerprint distribution matching across different environments. 
% for robust location predictions. 
During dynamic adversarial learning, the \textit{global domain discriminator} and the \textit{local subdomain discriminators} are trained to differentiate source CSI samples from target data, and further
DadLoc can automatically weigh the relative importance of global and local distribution adaptations with the \textit{dynamics-adaptive factor} $\mu$. Dynamic adversarial adaptation is carried out to enhance the environmental adaptability of the proposed DadLoc system effectively.

%environmental dynamics-resistant representations are learned with the feature extractor and the location label predictor that confuse the global and local domain discriminators.  
%Furthermore, due to the uncertainty of the location prediction in harsh multipath environments, we adopt an entropy-aware metric $E$ to hold the uncertainty of location label prediction and then condition the domain discriminator for effective and safe transfer. 
%Finally, 

\begin{figure}[tbp]
\centering
\includegraphics[width=0.35\textwidth]{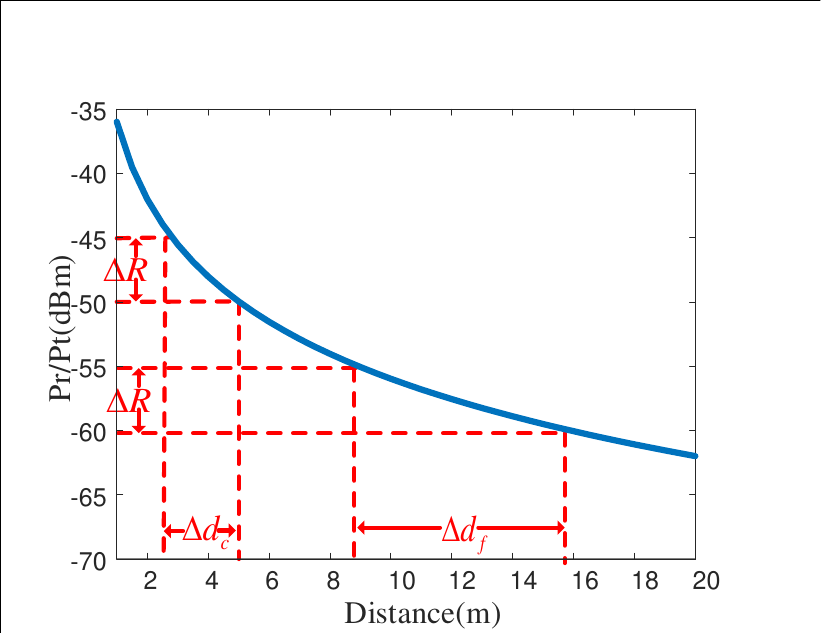}
\caption{Location Uncertainty }
\label{fig_pl}
\end{figure}

Furthermore, DadLoc incorporates dynamic adversarial adaptation network with \textit{prediction uncertainty suppression} to gain the effectiveness of accuracy improvement.  
We investigate the root cause of coarse localization accuracy which mainly involves the representation capability of location fingerprints. 
According to the path loss model \cite{LPLM} as the illustration  
 in Fig. \ref{fig_pl}, RF received strength decays logarithmically with the propagation distance. 
 An identical RF fading $\Delta \mathtt{RSS}$ can involve a smaller distance variance $\Delta d_c$ at closer locations to the transmitter, or a larger coverage  $\Delta d_f$ at faraway locations with the uncertainty of location information. 
RF fingerprints at  different locations have different capabilities of both signal representation and  location discrimination  which  are further entangled with the environmental dynamics \cite{LSTMLoc,MitLoc}. 
In this case, it is highly risky to learn transferable representation with the uncertainty of the discriminative information \cite{TransDis, CDAN, DDAN}, which cannot guarantee that the distribution discrepancy  across domains is sufficiently small with the equal distribution of source  and target data, even if the domain discriminator is fully confused.  Toward effective and safe transfer, 
DadLoc quantitatively evaluates the location uncertainty of RF fingerprints and removes this negative impact on dynamic adversarial adaptation to learn finer transferable representation for yielding remarkable performance gains.

%which attempts to quantify and enhance the location discriminability of the learned dynamic adaptation model for accuracy improvement gains. 

On receiving CSI queries, current feature representations are obtained by using the learned dynamic adversarial adaptation network, and the location estimation is worked out with the optimal matching of the location label predictor. DadLoc will automatically accommodate the new environmental change and perform online location estimations. 

\subsection{Adversarial Adaptation Network on Location Prediction}

%\subsection{Location Prediction Model}
%DadLoc  first builds up the  location prediction model with three components   of  DNN as \textit{radio image construction}, the \textit{feature extractor}, and the \textit{location label predictor}. The detailed design of the location prediction model is  described as follows.
An adversarial adaptation network \cite{DATN,DDAN} is adopted to achieve location predictions with the environmental dynamics, and the detailed design and implementation are presented as follows. 

%\textit{1)\; Radio Image Construction}
 
%Recent 
% for radio image construction based on the spatial correlation of the adjacent pixels in visual images.
 %WiFi-based  localization schemes \cite{deeploc, ConFi} prefer to transform  CSI values into radio images, since  commercial WiFi devices have been equipped with multiple antennas with the modulation of orthogonal frequency division multiplexing (OFDM) \cite{tool2011}, leading to CSI image with the spatial correlation of the adjacent pixels \cite{ILOT}. Furthermore, 
 
 Inspired by the strong representation capability of CNN to effectively gain accuracy improvement, advanced
works  \cite{deeploc, ITOL} prefer to transform  CSI values into radio images which contain  the spatial correlation of the adjacent pixels \cite{ILOT}, due to CSI values at multiple antennas and multiple subcarriers with the modulation of orthogonal frequency division multiplexing (OFDM) \cite{tool2011}. 
For radio image construction, we collect  $K$ vectors of CSI values as a  RF frame in a short time to construct a radio image  on the $m$th antenna with $N$ subcarriers  as 
\begin{equation}
   \mathbf{x}^m =
    \begin{bmatrix}
    x_{11}&x_{12}&\cdots&x_{1N}\\
    x_{21}&x_{22}&\cdots&x_{2N}\\
    \vdots&\vdots&\ddots&\vdots\\
    x_{K1}&x_{K2}&\cdots&x_{KN}\\
    \end{bmatrix}.
\end{equation}
The antennas on a WiFi device are considered as RF channels of a radio image like the RGB channels of an image, usually $M=3$. The radio image $\mathbf{x}_i=\{\mathbf{x}_i^m\}_{m=1}^M$ can be constructed and attached with the corresponding location label $y_i$. The data collection of radio images at all predefined RP is composed of the source domain $\mathcal{S}$. In a new indoor environment, CSI values are recorded and transformed to radio images as the target domain $\mathcal{T}$ without the location labels. 
% With the labeled data, the radio images $\mathbf{x}$ own the corresponding location label $y\in \mathcal{Y}$ which is the set of all locations in the interested area. The source domain denotes $\mathcal{D}_{s}=\left\{\left(x_{i}^{s}, y_{i}^{s}\right)\right\}_{i=1}^{n_{s}}$, and the target domain without the location labels $\mathcal{D}_{t}=\left\{x_{i}^{t}\right\}_{i=1}^{n_t}$  from the new environment. 

We design convolution neural networks (CNN) to learn feature representation  which can characterize  certain spatial relationships underlying radio images. %which is also informative to learn environment-independent features with dynamic adversarial learning in our scheme. 
%\textit{2)\; Feature Extractor}
%%\subsubsection{Feature Extractor}
%
%In the feature extractor, we design the CNN architecture to learn feature representation $\boldsymbol{f}$ as 
%\begin{equation}
%    \boldsymbol{f} = G_f(\mathbf{x};\theta_f),
%\end{equation}
%where $G_f(\cdot;\theta_f)$ is the function of the feature extractor with the learnable parameters $\theta_f$. 
%Specifically, our 
The feature extractor  $F$ in the CNN architecture consists of the convolutional layers, the max-pooling layers, and the fully connected (FC) layer. 
%, which are explicitly illustrated as DadLoc's network architecture in Fig. 3. 
With respect to deep adaptation networks, since the domain-specific features  linger along the deep layer with high probabilities \cite{DNN2014}, DadLoc  can learn informative feature representation with  multiple convolutional layers.
% and further attempts to reduce their distribution differences by dynamic adversarial learning. 
%% \begin{figure}[tbp]
%% 	\centering
%%  \includegraphics[width=0.22\textwidth]{DanLoc/feature.pdf}
%% 	\caption{CNN Architecture.}
%% 	%\label{fig_sa}
%% \end{figure}
%
%%\subsubsection{Location label Predictor}
%
%\textit{3)\; Location Label Predictor}
%
%After  the CNN module in the DadLoc system, 
The deep features $\boldsymbol{f}$ are fed to the fully connected  layers for the location label prediction.
The location label predictor $G$ is trained with source fingerprints, and the loss of the location label predictor is expressed as
\begin{equation}
   \mathcal{L}_{y}=
   -\frac{1}{n_{s}} \sum_{\mathbf{x}_{i} \in \mathcal{S}} \log G\left(F\left(\mathbf{x}_{i}\right), y_i\right).
\end{equation}
% learns a function $G$ which is parameterized as 
% \begin{equation}
%    G( F(\mathbf{x});\mathbf{V},\mathbf{c})=\operatorname{softmax}\left(\mathbf{V}G_f(\mathbf{x})+\mathbf{c}\right). 
% \end{equation}
%Therefore,  
%the loss  of the location label predictor is formulated with the negative log-probability as 
%\begin{equation}
%    \mathcal{L}_{y}=
%    %\frac{1}{n_{s}} \sum_{\mathbf{x}_{i} \in \mathcal{S}} \mathcal{L}_r \left( G_{y}\left(G_{f}\left(\mathbf{x}_{i}\right)\right),y_i\right),
%    -\frac{1}{n_{s}} \sum_{\mathbf{x}_{i} \in \mathcal{S}} \log G_{y}\left(G_{f}\left(\mathbf{x}_{i}\right);\theta_y\right)_{y_i}.
%\end{equation}
%%where $\mathcal{L}_r$ is the cross-entropy function of the location labels and $R$ is the number of RPs in the interested environment.
%where $\theta_y=\{\mathbf{V},\mathbf{c}\}$ is the parameters of $G_y$. 
%DadLoc can determine  location estimations with the location label prediction which involves the discriminative information underlying the training features,   so that the output of the location label prediction is also informative to learn environment-invariant features for robust localization. 
% In order to realize robust fingerprint-based localization with diverse environmental changes, DaLoc proposes self-adaptive adversarial learning to train dynamics-representations with the feature extractor and the location label predictor as follows.
%In order to enable this location predictor to adapt the environmental changes, the distribution shifts  should be eliminated with the transferability and discriminability

During domain adversarial learning,  the distribution shifts across source and target environments are reduced to learn transferable feature representation with the minimax game that 
\textit{the domain discriminator} $D$ is trained to distinguish the feature representation of the source samples from the target data, whereas the feature extractor   $F$ attempts to confuse the domain discriminator.   Meanwhile, the deep adaptation network is optimized to minimize the error of the location label predictor $G$ on the source fingerprints. Specifically,  the feature extractor $F$ is trained by maximizing the loss of domain discriminator $D$, while the domain discriminator $D$ is trained by minimizing its loss.
The adversarial  adaptation model is formulated as
\begin{equation}
	\begin{split}
	M_0\left(\theta_{f}, \theta_{y}, \theta_{d}\right)  =&%\frac{1}{n_{s}}
	\mathbb{E}_{\mathbf{x}\in \mathcal{S}} \left[ \mathcal{L}_{y}\left(G\left(F\left(\mathbf{x}\right)\right), y\right) \right]\\
		& -\lambda %\frac{\gamma}{n_{s}+n_{t}} \sum
		\mathbb{E}_{\mathbf{x} \in \mathcal{S} \cup \mathcal{T}}\left[\mathcal{L}_{d}\left(D\left(F\left(\mathbf{x}\right)\right), d\right)\right],
	\end{split}
\end{equation}
where  $\lambda$ is a trade-off parameter to the penalty of adversarial domain adaptation, 
%\begin{equation}
%\begin{align}
%%	&\min _{\theta_{f}, \theta_{y}} \max _{\theta_{d}} \mathcal{L}_{y}\left(\theta_{f}, \theta_{y}\right)+\gamma \mathcal{L}_{adv}\left(\theta_{f}, \theta_{d}\right), \\
%	&\mathcal{L}_{adv}\left(\theta_{f}, \theta_{d}\right)=\notag\\
%	&\frac{1}{n_{s}} \sum_{i=1}^{n_{s}} \log \left[D\left(F\left(\mathbf{x}_{i}^{s}\right)\right)\right]+\frac{1}{n_{t}} \sum_{j=1}^{n_{t}} \log \left[1-D\left(F\left(\mathbf{x}_{j}^{t}\right)\right)\right], \\
%	&\mathcal{L}_{y}\left(\theta_{f}, \theta_{y}\right)=\frac{1}{n_{s}} \sum_{i=1}^{n_{s}} \ell \left(G\left(F\left(\mathbf{x}_{i}^{s}\right)\right), y_{i}^{s}\right),
%\end{align}
%\end{equation}
%  $\mathcal{L}_y$ is the loss of the location label predictor as
%  \begin{equation}
%    \mathcal{L}_{y}(\theta_f,\theta_g)=
%    -\frac{1}{n_{s}} \sum_{\mathbf{x}_{i} \in \mathcal{S}} \log G\left(F\left(\mathbf{x}_{i}\right), y_i\right),
% \end{equation}
 and $\mathcal{L}_d$ is the cross-entropy loss of the domain  discriminator as 
 \begin{equation}
%\begin{split}
\mathcal{L}_{d}= 
\frac{1}{n_s}\sum_{i=1}^{n_s}\log {D\left(F\left(\mathbf{x}_{i}\right)\right)}+\frac{1}{n_t}\sum_{i=1}^{n_t} \log \left[1-D\left(F\left(\mathbf{x}_{i}\right)\right)\right], 
%\end{split}
\end{equation}
 $\theta_{f}$, $\theta_{d}$, and $\theta_{y}$ are the learnable parameters of  the feature extractor, the domain discriminator, and the location label predictor, respectively. 
With the mini-max game, the parameters will be obtained with the training convergence as
\begin{equation}
	\begin{split}
		&\left(\hat{\theta}_f, \hat{\theta}_y\right)=\min_{\theta_{f}, \theta_{y}} M_0\left(\theta_{f}, \theta_{y}, \theta_{d}\right)\\
		&\left(\hat{\theta}_{d}\right)=\max_{\theta_{d}} M_0\left(\theta_{f}, \theta_{y}, \theta_{d}\right).
	\end{split}
\end{equation}

While adversarial adaptation networks are adopted to achieve robust location estimations with the environmental dynamics \cite{DRF,ILOT}, the feature confusion with sparse spatial constraint leads to the degeneration of  the location discriminability \cite{TransDis} with coarse localization. Therefore, DadLoc develops dynamic adversarial adaptation network to learn finer transferable feature representation for robust fingerprint-based localization.

%\subsubsection{Global Domain Discriminator}
\subsection{Dynamic Adversarial Adaptation Network on Automatic Radio Map Adaptation}

In order to accommodate complicated environmental changes, DadLoc further performs a finer-grained distribution adaptation  with the distribution matching of both features and fingerprints in a dynamics-adaptive manner.  
%further reduces the fingerprint shifts with local subdomain adaptation, 
Specifically, the feature distribution adaptation can follow the  adversarial adaptation network with
the objective of global distribution adaptation is expressed as 
\begin{equation}
	\mathcal{L}_{g}=\mathbb{E}_{\mathbf{x} \in \mathcal{S} \bigcup \mathcal{T}} \left[\mathcal{L}_d \left(D\left(F\left(\mathbf{x}\right)\right),d\right)\right] .
\end{equation}
After that, 
we split the domain discriminator $D$ into location-label-wise domain discriminator $D^r, r=1,2,\ldots,R$, where each is also responsible for matching fingerprint distribution, as the features between the source and target data associated with the location label $r$. 
With the location label prediction, the output of the location label predictor $\hat{y}^r$ is a probability distribution over the location label space which indicates the probability of associating the extracted features with each RP. 
DadLoc further performs fingerprint distribution adaptation  which relies on conditional/local distribution  matching cross domains with the corresponding elements of $\hat{y}_i^r$.
The loss function of the local subdomain discriminator is formulated as  
\begin{equation}
	\mathcal{L}_{l}= \sum_{r=1}^{R} \mathbb{E}_{\mathbf{x} \in \mathcal{S}\bigcup \mathcal{T}}\left[\mathcal{L}_{d} \left(D^{r}\left(\hat{y}^r F\left(\mathbf{x}\right)\right),d\right)\right].
\end{equation}
%where $d_i^r$ is the domain label of the input sample $\mathbf{x}_i$ at the $r$th RP. 
%If $d=0$, the input sample comes from the source data, or such label is equal to 1 that the input sample is the target data. 

Furthermore, DadLoc evaluates the importance of both global and local distribution adaptation in a dynamic self-adaptive manner. We denote the  dynamics-adaptive factor $\mu$ to control such dynamic distribution adaptation  for automatically adapting to diverse environmental changes, which can be formulated as
\begin{equation}
	\mathcal{L}_{adv}^{dy}=(1-\mu)\mathcal{L}_{g}+\mu \mathcal{L}_{l},
\end{equation}
where   the value of $\mu$ ranges from 0 to 1 with the surrounding environmental changes. When $\mu$ is prone to 0, the global distribution adaptation dominates to enable automatic radio map adaptation, otherwise, that is the local distribution adaptation with the environmental dynamics.

% learns environment-invariant features through 
%the dynamic domain adversarial networks  which incorporate the location prediction model with dynamic adversarial learning to perform finer-grained distribution alignment across domain in different environments. The dynamic domain adversarial networks contain the \textit{global domain discriminator} $G_{d}$, the \textit{local domain discriminator} $G_d^r$ with uncertainty-aware balancing, and the dynamic-adaptive factor. They are based on co-training strategy as follow. 

\begin{figure*}[tbp]
	\centering
	\includegraphics[width=0.95\textwidth]{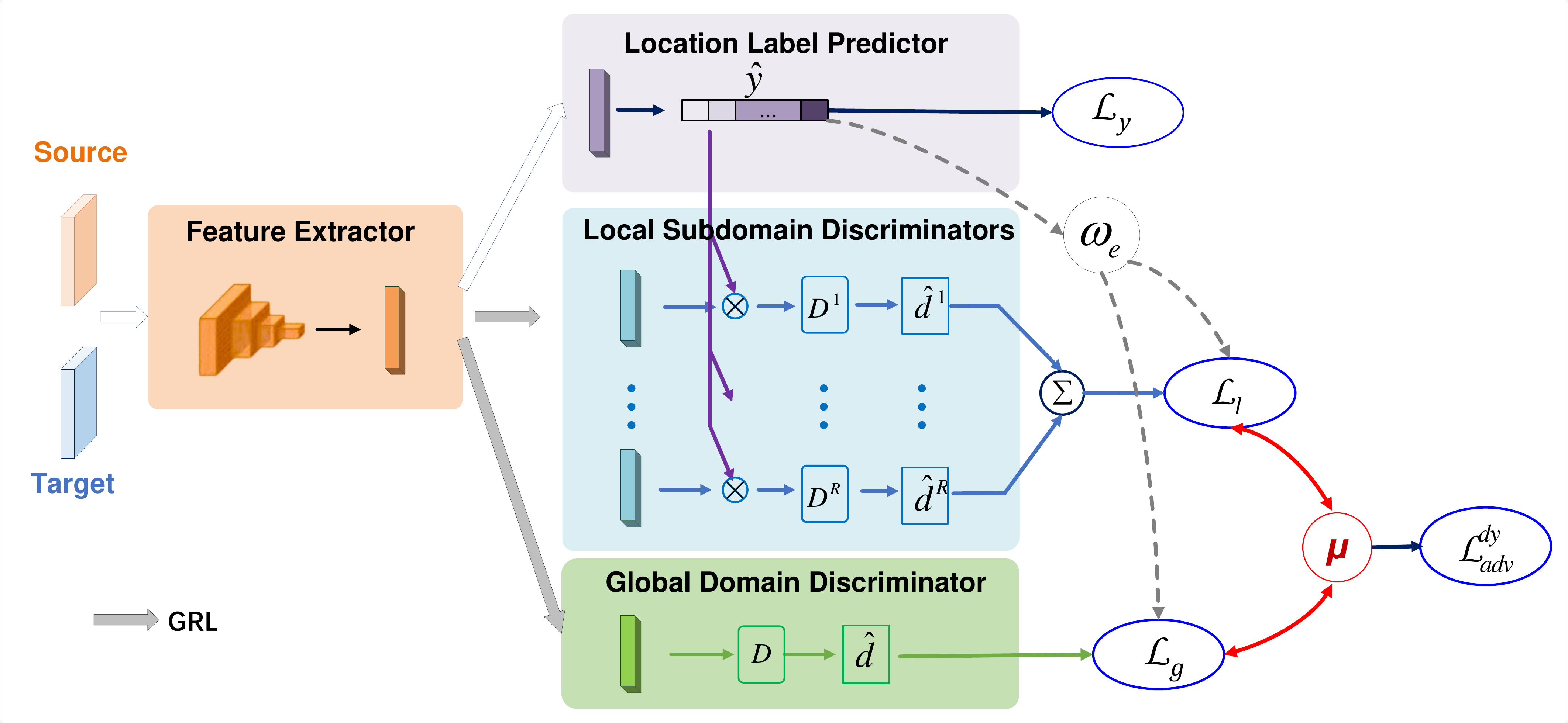}
	\caption{DadLoc's Architecture for dynamic distribution adaptation with the feature extractor, the location label predictor, and global and local domain discriminators, where $\hat{y}$ is the predicted location label, $\mathcal{L}_y$ is the loss of the location label predictor, and $\hat{d}$ and $\hat{d}^r$ are the predicted domain labels, $\mathcal{L}_g$ and $\mathcal{L}_l$ is the loss of the global and local domain discriminators, respectively; $\mu$ is the dynamics-adaptive factor; $\omega_e$ is the uncertainty-aware weights of the location label prediction.    }
	\label{fig_sa}
\end{figure*}
With the developed DAAN, the learned deep representations are used to learn and update $\mu$, and then 
 DadLoc can directly rely on the losses of these discriminators to automatically up-weight the  dynamics-adaptive factor by reducing dynamic  distribution discrepancy across environments.  
We resort to $\mathcal{A}$-distances to measure the global and local distribution discrepancies across different environments, and further the weighted importance of the global and local distribution minimization is evaluated by the dynamics-adaptive factor $\mu$. 
The dynamic distribution  discrepancy $\mathtt{Disc}(\mathcal{S}, \mathcal{T})$ is defined with the $\mathcal{A}$-distance of the global and local  distributions as
\begin{equation}
	\mathtt{Disc}(\mathcal{S}, \mathcal{T})=(1-\mu)\mathcal{A}_g\left(\mathcal{S}, \mathcal{T}\right)+\mu \sum_{r=1}^{R}\mathcal{A}_l\left(\mathcal{S}^{(r)}, \mathcal{T}^{(r)}\right),
\end{equation}
where $\mathcal{A}_g$ and $\mathcal{A}_l$ are denoted as the global $\mathcal{A}$-distance and the local $\mathcal{A}$-distance, respectively.  During an adversarial learning of dynamic adaptation network, DadLoc 
can calculate $\mathcal{A}_g$ with the loss of global domain discriminator as follow
\begin{equation}
	\mathcal{A}_g(\mathcal{S}, \mathcal{T})=2(1-2\mathcal{L}_g),
\end{equation}
and the local $\mathcal{A}$-distance is figured out as
\begin{equation}
\sum_{r=1}^{R}\mathcal{A}_l\left(\mathcal{S}^{(r)},  \mathcal{T}^{(r)}\right)=2(1-2\mathcal{L}_l).
\end{equation}
Therefore, the dynamics-adaptive factor $\mu$ is estimated as
\begin{equation}
	\hat{\mu}=\frac{\mathcal{A}_g\left(\mathcal{S}, \mathcal{T}\right)}{\mathcal{A}_g\left(\mathcal{S}, \mathcal{T}\right)+\frac{1}{R} \sum_{r=1}^{R} \mathcal{A}_l\left(\mathcal{S}^{(r)}, \mathcal{T}^{(r)}\right)}.
\end{equation}
Within the developed DAAN, the value of the dynamics-adaptive factor $\mu$ is obtained after each epoch of iteration. 
DadLoc can be automatically updated by  minimizing the losses of both global and local domain discriminators to learn a robust $\mu$ until the training convergence.

\subsection{Prediction Uncertainty Suppression}
%\textit{3)\;Uncertainty-Aware  Balancing}
As aforementioned introduction, different fingerprints have different capabilities of feature representation and location discrimination which further leads to the uncertainty of the location prediction. 
% According to LDPL model, RF received strength decays logarithmically with propagation distance, and thus, an identical   RF fading can involve a smaller distance variance at closer locations to AP, or a larger coverage at faraway locations. 
%CSI fingerprints from the closed AP  have relative stronger discriminability than that of faraway AP, which le
%can indicate the location prediction with lower uncertainty. 
DadLoc expects that the strong discriminability of CSI fingerprints with lower location uncertainty is emphasized whereas high location uncertainty from faraway fingerprints is weakened, during dynamic adversarial learning with the developed DAAN.
%when the matching of fingerprint distributions is performed with the environmental changes. 
Thus, we quantitatively evaluate the location uncertainty of fingerprint representation via the information entropy as
\begin{equation}
    H(y)=-\sum_{r=1}^R \hat{y}^r\log \hat{y}^r,
\end{equation}
where $\hat{y} =G(F(\mathbf{x})) $ is the softmax probability of the location label predictor. DadLoc  further imposes the effectiveness of transferable representation in the developed DAAN  with an uncertainty-aware weight \cite{CDAN} as
\begin{equation}
	\mathcal{L}_{adv}^{unc}= \sum_{r=1}^{R} \mathbb{E}_{\mathbf{x} \in \mathcal{S}\bigcup \mathcal{T}}\left[\omega_e(\mathbf{x})\mathcal{L}_{d} \left(D^{r}\left(\hat{y}^r F\left(\mathbf{x}\right)\right),d\right)\right], 
\end{equation}
where 
\begin{equation}
\omega_e(\mathbf{x})=1+\exp{\left(-H(G(\mathbf{x}))\right)}.
\end{equation}
At the same time, the uncertainty-aware weight in the feature level is also calculated to condition the matching of feature distribution as
\begin{equation}
\mathcal{L}_{g}^{unc}=\mathbb{E}_{\mathbf{x} \in \mathcal{S} \bigcup \mathcal{T}} \left[\omega_e(\mathbf{x}) \mathcal{L}_d \left(D\left(F\left(\mathbf{x}\right)\right),d\right)\right].
\end{equation}
The functional objective of our improved dynamics-adversarial-learning is reformulated with the uncertainty-aware regularization as
\begin{equation}
\mathcal{L}_{adv}^{dy}=(1-\mu)\mathcal{L}_{g}^{unc}+\mu \mathcal{L}_{adv}^{unc}.
\end{equation}

\subsection{DadLoc's Learning Procedure}
Once the developed DAAN is trained to eliminate global and local distribution shifts,  dynamic distribution adaptation degenerates to the location predictions of target data with semi-supervised learning. Due to the unlabeled target data, it is significant to preserve the structure for effective transfer.  DadLoc enables the location predictor trained with source fingerprints to be directly utilized for target data. 
Under the entropy minimization principle \cite{EMP}, it encourages all the unlabeled target data to own highly confident predictions with the following term of the conditional entropy as
\begin{equation}
\mathcal{L}_{tar}^{unc}=\mathbb{E}_{\mathbf{x} \in \mathcal{T}} H(G(F(\mathbf{x}))).
\end{equation}

We finally integrate source-supervised training, target-unsupervised training, and source-target dynamic adversarial adaptation  to learn  finer transferable representation for robust localization. By incorporating Eq. (3), (18), and (17),  
the objective of the proposed DadLoc is expressed as
\begin{equation}
	\mathcal{L}=\mathcal{L}_y+ \gamma\mathcal{L}_{tar}^{unc}-\lambda \mathcal{L}_{adv}^{dy},
\end{equation}
where $\gamma$ and $\lambda$ are empirical trade-off parameters. At the learning time, DadLoc works out the parameters $\theta_f$ of the feature representation to learn  feature mapping that maximizes the losses of both global and local domain discriminators for the confusion of feature and fingerprint distribution, while the parameters of these discriminators $\theta_d$ and $\theta_d^r$ minimize their losses. At the same time, the loss of the location label predictor is minimized with source fingerprints. 
The  min-max  optimization of the proposed DadLoc is to find out the optimal parameters  $\hat{\theta}_f$, $\hat{\theta}_y$, $\hat{\theta}_d$, and $\hat{\theta}_d^r$ as
\begin{equation}
\begin{split}
		&\left(\hat{\theta}_f, \hat{\theta}_y\right)=\min_{\theta_{f}, \theta_{y}} M\left(\theta_{f}, \theta_{y}, \theta_{d},\theta_d^r\right)\\
		&\left(\hat{\theta}_{d}, \hat{\theta}_d^r|_{r=1}^R\right)=\max_{\theta_{d},\theta_d^r} M\left(\theta_{f}, \theta_{y}, \theta_{d},\theta_d^r\right).
\end{split}
\end{equation}

\section{Experimental Evaluation}
In this section, we introduce the experiment methodology and the performance evaluation  of the proposed DadLoc system. We also prototype the DadLoc system in real indoor scenarios using commercial WiFi devices and conduct extensive experiments in diverse  changing environments. Moreover, we continuously evaluate the localization performance for more than 8 months.  

\begin{figure*}[thbp]
	\centering
	\subfigure[Hall]{
		\includegraphics[width=0.22\textwidth]{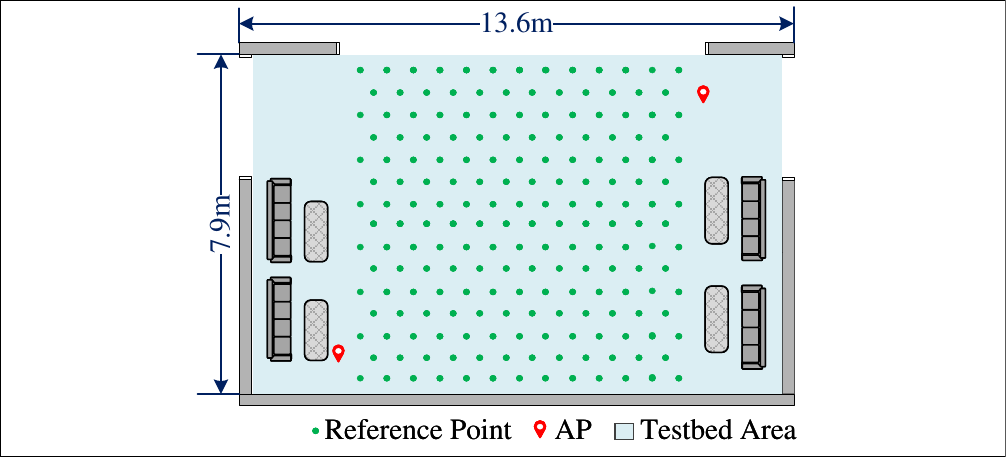}} \quad 
	\subfigure[Corridor]{
		\includegraphics[width=0.35\textwidth]{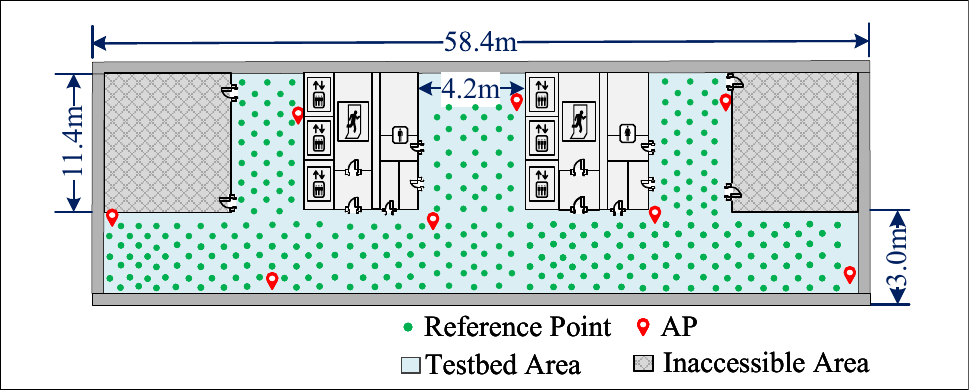}} \quad 
	\subfigure[Lounge]{
		\includegraphics[width=0.32\textwidth]{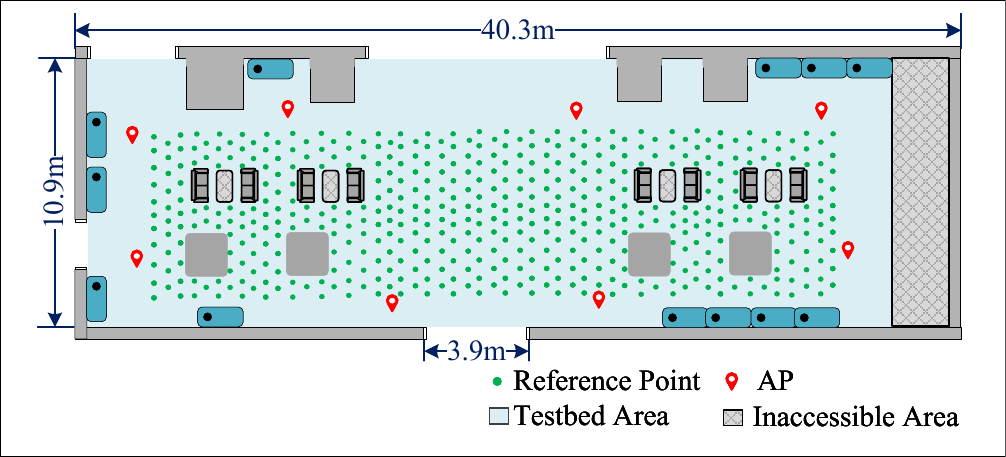}}
	\caption{Floor layouts of indoor scenarios. }
	\label{fig_floorplan}
\end{figure*}

\subsection{Experiment Methodology}

\subsubsection{ Data Collection}

We collect the  CSI fingerprints at every RPs by using commercial WiFi infrastructures  which are equipped with off-the-shelf  802.11n 5300 NICs. The injection mode is mounted on the transceivers with MIMO-OFDM modulation scheme  in 5GHz WiFi spectrum by the sampling frequency of 100Hz \cite{tool2011}. In typical indoor environments, we have provided the localization service of the proposed DadLoc for over 8 months. On the initial radio map construction, we record CSI readings  in the experimental areas. After that, we conduct CSI fingerprint collection with the site survey every three days for two weeks. For the performance evaluation of the long-term environmental changes, we repeat the site survey procedure with CSI fingerprints lasting 6 months.

We deploy the proposed DadLoc  in three typical indoor scenarios with   different RF propagation conditions and the respective floor layouts illustrated in Fig. \ref{fig_floorplan}. 
Specifically,   the \textit{hall} is the size of 98 $m^2$   with  more light of sight (LOS) propagation in Fig. \ref{fig_floorplan}(a). We set 188 RPs with the adjacent spacing of 0.8m and there are 187 testing locations with the corresponding CSI queries for the performance evaluation of the proposed  DadLoc scheme. We place two WiFi receivers to record CSI measurements. The \textit{corridor } is the dense multipath scenario with 360 RPs and the same number of test locations, in the size of 298 $m^2$ with daily crowded users as the illustration of Fig. \ref{fig_floorplan}(b). In the whole area, eight APs are set to collect CSI readings at each RP. 
The \textit{lounge}, as the third indoor scenario, has furniture and obstacles with non-light-of-sight (NLOS) wireless propagation. 425 RPs are deployed in this complicated environment, and 426 test points are chosen to estimate their locations. There are also eight APs in the lounge.

%\begin{figure}[thbp]
%	\centering
% \includegraphics[width=0.48\textwidth]{DadLoc_NA.pdf}
%	\caption{Network Architecture}
%	\label{fig_pus}
%\end{figure}

\subsubsection{Network Architectures and Implementation Details}
The  architecture of the proposed DadLoc  is illustrated in Fig. \ref{fig_sa}. DadLoc has the backbone structure for all the scenarios. The different sizes of the input data lie in the number of RPs in various indoor interested environments. We first use source CSI fingerprints to train the feature extractor and the location label predictor with the source fingerprints, and then fine-tune them through dynamic  adversarial learning with both global and local domain discriminators in a dynamics-adaptive manner. %After that, the training model is evaluated on the validation set to avoid overfitting. 

In DadLoc network architectures, the feature extractor is composed of four CNN layers where each CNN layer includes two convolutional layers with Leaky-ReLu activation ($\alpha = 0.1$), one max-pooling, and mini-batch normalization with a batch size of 64. 
The location label predictor and the global and local domain discriminators are three fully connected (FC) layers, respectively. 
With respect to unsupervised domain adaptation tasks, all convolutional and pooling layers, and  the discrimination layer are fine-tuned via backpropagation with gradient reversal layer (GRL) \cite{DTN,DAAN}. Since the location label predictor is trained from scratch, its learning rate is set to be 10 times that of the other layers.
The mini-batch SGD optimization is conducted with a momentum of 0.9. 
The learning rate is adjusted during the  SGD with the formula \cite{DATN}
as $\eta_n=\eta_{0}/(1+\alpha \cdot n)^{\beta}$ where $n$ is the training progress linearly ranging from 0 to 1, $\eta_0=0.01$, $\alpha=10$, and $\beta=0.75$. We fix $\lambda=1$ to update the domain discriminators within the developed DAAN.

\subsubsection{Comparative Approaches}
In order to  validate the extensive effectiveness of the proposed DadLoc, 
we compare its localization performance with classical deep-learning-based localization schemes:
\begin{itemize}
\item 
Supervised learning-based methods: WiDeep \cite{WiDeep},  CNNLoc \cite{CNNLoc}, and LESS \cite{l2l}.
%which adopts few-shot relation learning to characterize the proximity relationships among the neighboring fingerprints for robust localization. 
\item Unsupervised learning-based method: DeepFi \cite{DeepFi}. 
\item 
Domain-adversarial learning methods: iToLoc \cite{ILOT} and TransLoc \cite{DRF}. 
\end{itemize}
% and  DeepFi \cite{DeepFi} with unsupervised learning-based method, DeepFi \cite{DeepFi}. Meanwhile, we further evaluate the environmental adaptability of the proposed DadLoc with the comparison of LESS \cite{l2l} which adopts few-shot relation learning to characterize the proximity relationships among the neighboring fingerprints for robust localization. 
% Furthermore, 
% by using domain-adversarial learning, we evaluate the system performance of DadLoc as well as advanced approaches iToLoc \cite{ILOT} and TransLoc \cite{DRF} to evaluate the accuracy improvement of robust localization.

For the fairness of performance comparison, we use the same CNN architecture to train deep feature representations with the input radio images for all robust localization solutions, such as CNNLoc, iToLoc, TransLoc, and LESS.

%The localization errors  with different algorithms are summarized in TABLE I with the regularization parameters as $\lambda=10$, $\gamma=1$, $\eta=0.1$, and $p=5$.  

% \subsubsection{Evaluation Method} 

% To validate the effectiveness of the proposed approach on robust localization,
% we conduct extensive experiments for  performance evaluation with diverse environmental changes.  
% Tasks on location estimations come from Environment 1 to Environment 2 (Env. 1 $\to$ Env. 2) in a short interval of  one week, and  then the sudden environmental changes are caused by  user motion in Environment 3. 
% The long-term changing fingerprints are  collected from Environment 3 to Environment 4 (Env. 3 $\to$ Env. 4) for more than 6 months with the crowded users.  
% We calculate the key evaluation indicators of regression, such as  root mean square error (RMSE) $\hat{\epsilon}$, mean absolute errors (MAE) $\bar{\epsilon}$, and standard deviation (STD) $\sigma$, respectively. 

\subsection{Performance Evaluation and Discussion}
To validate the effectiveness of the proposed approach on robust localization, 
we extensively evaluate the localization performances of DadLoc and other comparative approaches as well. We further conduct the performance evaluation of their robustness  with diverse indoor environmental changes.  
With extensive experimental results, DadLoc outperforms the state-of-the-art  approaches to handle the specific tasks on target location prediction, especially  more complicated indoor scenarios. The performance analysis of the proposed DadLoc  in detail is as follows.

\subsubsection{Overall Performance Evaluation}
\begin{figure}[tbp]
	\centering
 \includegraphics[width=0.46\textwidth]{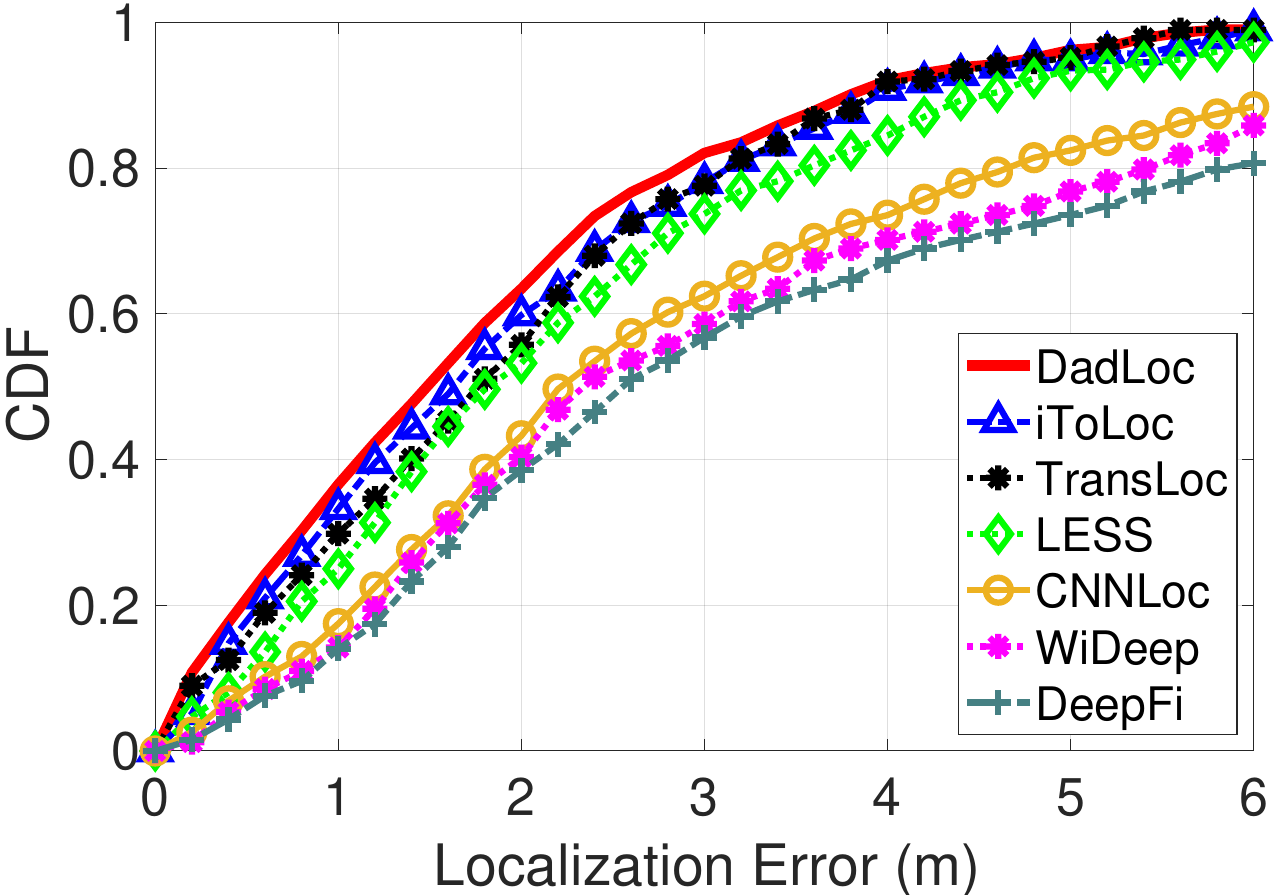}
	\caption{Overall Performance Comparison}
	\label{overall-cdf}
\end{figure}

We first demonstrate the overall performance of the proposed DadLoc system with all CSI measurements across 8 months. 
Fig. \ref{overall-cdf} illustrates the cumulative distribution functions (CDFs) of the location estimation errors which are the bias between the localization result and ground truth. 
Concerning adversarial-learning-based solutions, the proposed  DadLoc achieves the median localization error of $1.52m$ with various indoor environmental changes,  which attains the accuracy improvement of 9.2\% for iToLoc and   13.1\% for TransLoc, respectively.   The 90th percentile accuracy of the proposed DadLoc outperforms other systems by 19.1\% for iToLoc, and 25.6\% for TransLoc, respectively. Furthermore, DadLoc has gained an accuracy improvement of 16.5\% for LESS. With these localization results, the schemes on robust localization by incorporating more robust factors underlying CSI Fingerprints have shown superiority over a single one, such as TransLoc and LESS, and further their environmental adaptability 
are explicitly evaluated in the following subsection.

Additionally, although  CNN-based localization algorithm \cite{CNNLoc} can improve the accuracy of  the basic learning-based location solutions such as  WiDeep and DeepFi, they are difficult to  accurately estimate user locations with  indoor environmental dynamics. The median error of CNNLoc is $2.21m$ which leads to the limited accuracy for robust localization. WiDeep adopts denoising autoencoder to learn noise-tolerant features with an error of $2.32m$. DeepFi employs unsupervised learning to extract signal features by minimizing the reconstruction loss, which has a median error of $2.56m$.

The localization results of the proposed DadLoc system mainly benefit from dynamic distribution adaptation by automatically weighting the importance of both
global and local distribution adaptation.  With complicated indoor environmental changes for a long period, the proposed DadLoc can effectively learn finer transferable representation with the developed DAAN  and further achieve accuracy improvement for robust localization.

\subsubsection{Environmental Adaptability Evaluation}

We examine the robustness of the proposed  DadLoc system with diverse indoor environmental dynamics including the short-term localization performance mainly with unpredictable inference and the long-term localization performance with complicated and multivariate factors of the environmental dynamics.

\begin{figure}[tbp]
	\centering
	\subfigure[Short-Term Performance Comparison]{
		\includegraphics[width=0.4\textwidth]{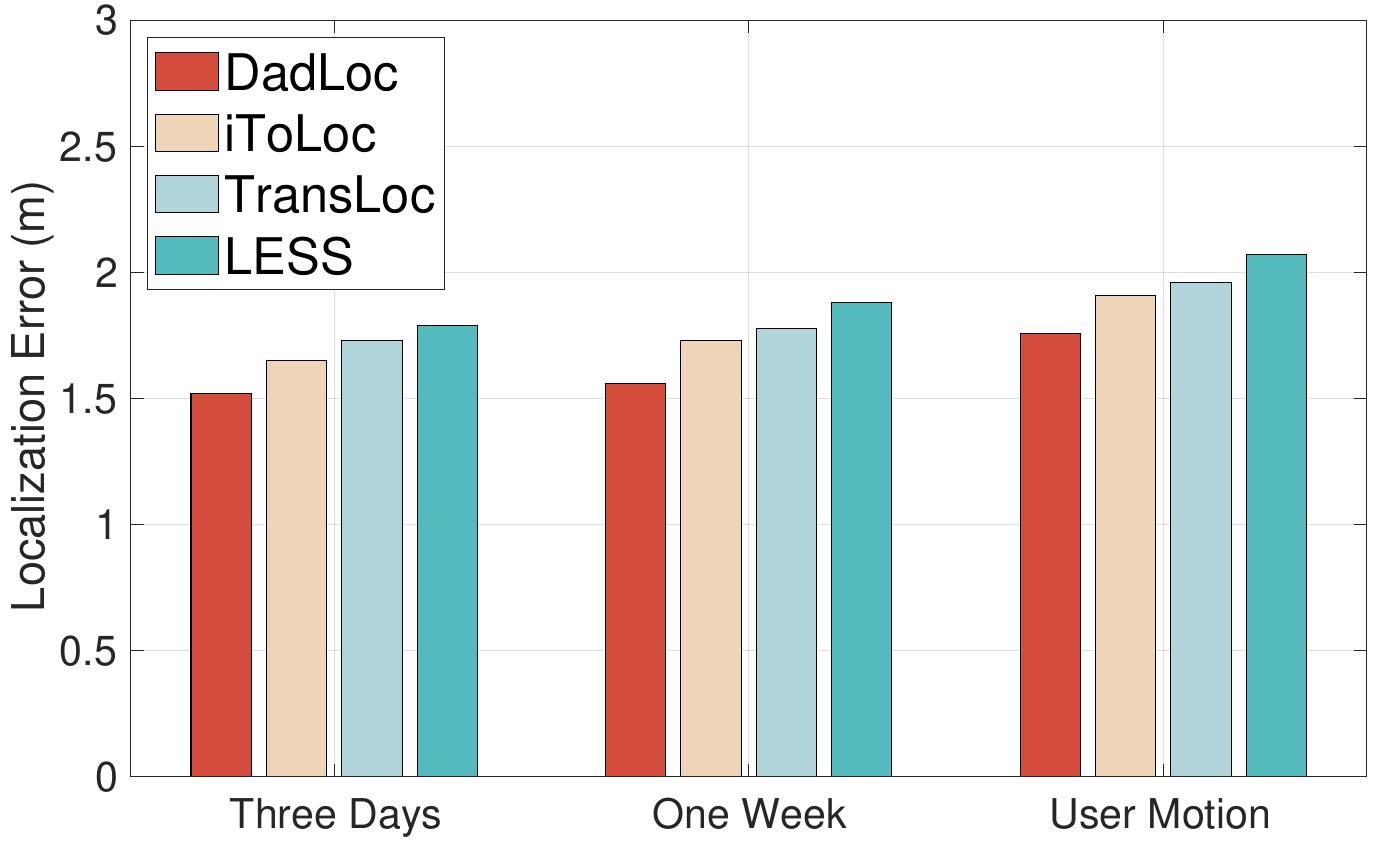}} \quad \quad
	\subfigure[Long-Term Performance Comparison]{
		\includegraphics[width=0.42\textwidth]{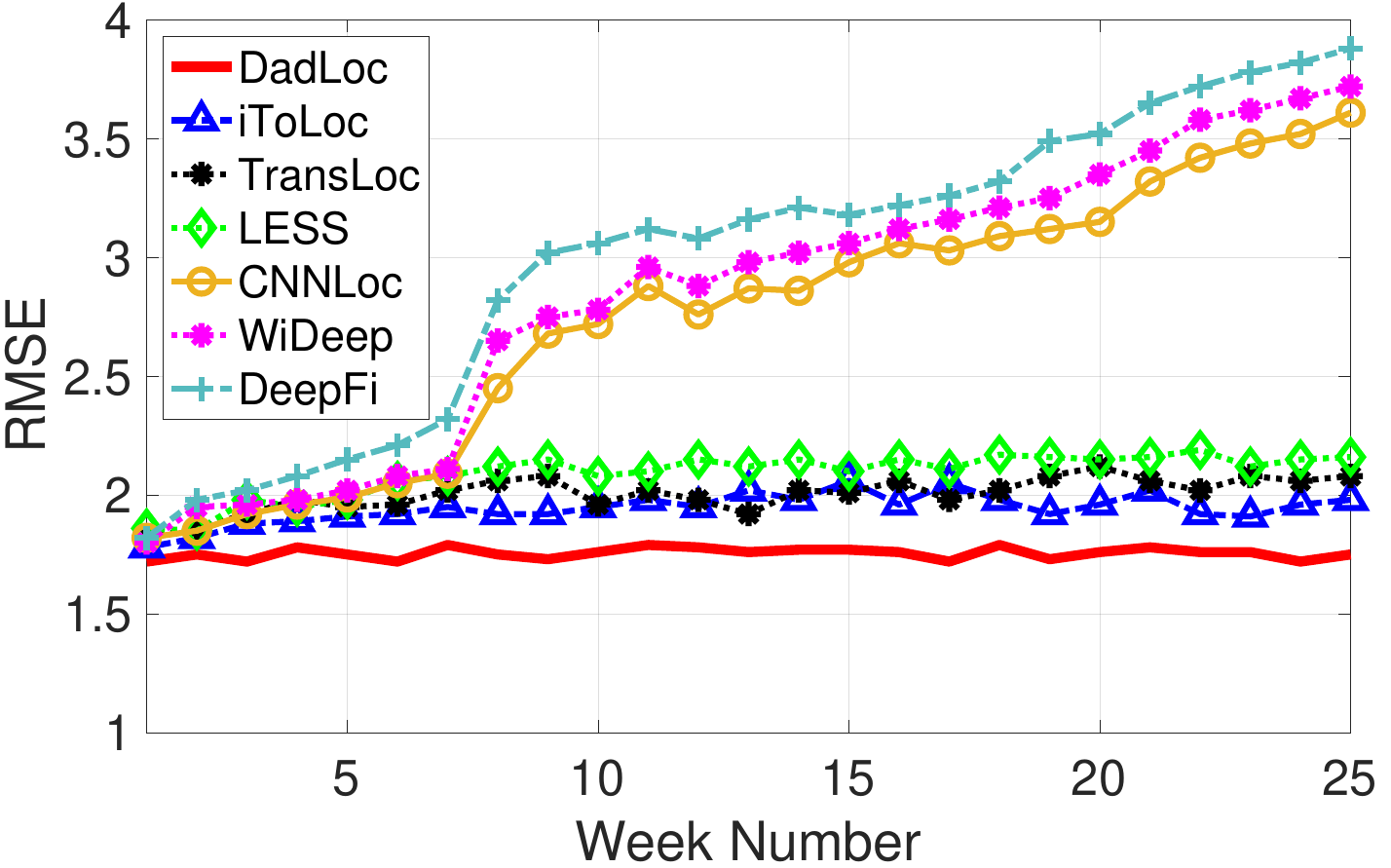}}
	\caption{Environmental Adaptability Comparison. }
	\label{fig_env}
\end{figure}

CSI fingerprints are recollected at different times such as on the third day and one week later with the crowded user motions. 
We evaluate the estimated errors of robust localization schemes such as DadLoc, iToLoc, TransLoc, and LESS for offering  a short-time LBS with indoor environmental changes. As the illustration of Fig. \ref{fig_env}(a), 
the localization error of the proposed DadLoc is $1.52m$ with  the  location accuracy improvements of iToLoc by 9.8\%, TransLoc by 12.1\%, and LESS by 13.9\%, on the third day, respectively.  
In one week, their respective localization errors are $1.56m$ for the proposed  DadLoc, $1.72m$ for iTocLoc, $1.78m$ for TransLoc, and $1.88m$ for LESS. 
With the crowded user motion in the tested areas, DadLoc achieves an average  accuracy of $1.76m$ which shows  accuracy improvement over others, such as $1.92 m$ for iToLoc, $1.98 m$ for TransLoc, and $2.07 m$ for LESS, respectively. These experimental results can verify that the proposed DadLoc system achieves better  localization performance with the robustness of unpredictable inferences in an indoor changing environment. 

%\subsubsection{Long-Term Performance Evaluation}
For long-term localization performance comparison, we continuously collect CSI fingerprints at different times for over 6 months. Fig. \ref{fig_env}(b) illustrates the accuracy tracking of every week.
The proposed DadLoc achieves an average error of $1.75m$ which performs more robust location estimations than other comparative methods for a long-running period in dynamic indoor environments, while there are complicated environmental dynamics in these indoor scenarios. 
With distribution adaptation based on adversarial learning, iToLoc and TransLoc have better accuracy than LESS for robust localization. 
%the CDF performance of real experiments  by providing location services with crowded users.
%The average error of the DadLoc system is $1.72m$ which obtains better location accuracy by 9.8\% of iToLoc and 13.6\% of TransLoc, respectively. 
In the eighth week, the deployment of furniture in the lounge has been changed.  Classical deep-learning-based schemes, such as CNNLoc, WiDeep, and DeepFi have extremely increasing errors with unpredictable environmental dynamics. However, DadLoc is negligibly impacted to achieve robust performance. 
The above localization results demonstrate that the proposed
DadLoc system mainly benefits from the developed DAAN to effectively enhance environmental adaptability and location discriminability of learned feature representation. 
%which effectively attains the improvements of adaptability and generalization capability on fingerprint-based localization. 

\subsubsection{Indoor Scenario Diversity}

We conduct the proposed approach with different conditions of wireless propagation in many typical indoor scenarios, such as the hall  with   LOS propagation, the corridor with  dense multipath environment, and  the lounge with NLOS propagation and crowded users. %The detailed performance of localization accuracy is shown in TABLE I with the comparison of deep-learning-based localization solutions. 
We calculate the key evaluation indicators of regression, such as  root mean square error (RMSE) $\hat{\epsilon}$, mean absolute errors (MAE) $\bar{\epsilon}$, and standard deviation (STD) $\sigma$, respectively. The performance analysis of the indoor scenario diversity is represented in the next subsection. 
%of the localization solutions based on deep domain adaptation as illustrated in Fig. \ref{indoor-scenario}, t
In the complicated indoor environment as the lounge,
the proposed DadLoc achieves the RMSE of $1.92m$ by 15.0\% for  iToLoc, 19.2\% for TransLoc,  and 23.8\% for LESS, respectively. 
In the corridor, the proposed DadLoc improves the accuracy by about 8.2\% for iToLoc,  9.1\% for TransLoc,  12.6\% for LESS, respectively. Meanwhile, the localization accuracy of the proposed DadLoc  in the hall outperforms by 8.7\% for iToLoc, 13.3\% for TransLoc, and 16.9\% for LESS, respectively. 
With these extensive experimental results, the proposed DadLoc scheme can achieve better localization performance than recent advances under more complicated RF  propagation which can obtain the robustness improvement of deep-learning-based localization with indoor environmental dynamics.

\subsection{Effectiveness  Analysis}

% \subsubsection{Effectiveness on Dynamics-adaptive Factor}
% We first investigate that it is necessary to conduct dynamic distribution adaptation for robust localization with complicated environmental changes. The developed DAAN is used to estimate the test locations in every indoor scenario by the constraint  $\mu \in \{0,0.1,\ldots,0.9,1\}$. We find out that the localization accuracy varies with different values of $\mu$ in the same indoor scenario.

\subsubsection{Effectiveness on Dynamic Distribution Adaptation}

\begin{figure}[thbp]
	\centering
 \includegraphics[width=0.4\textwidth]{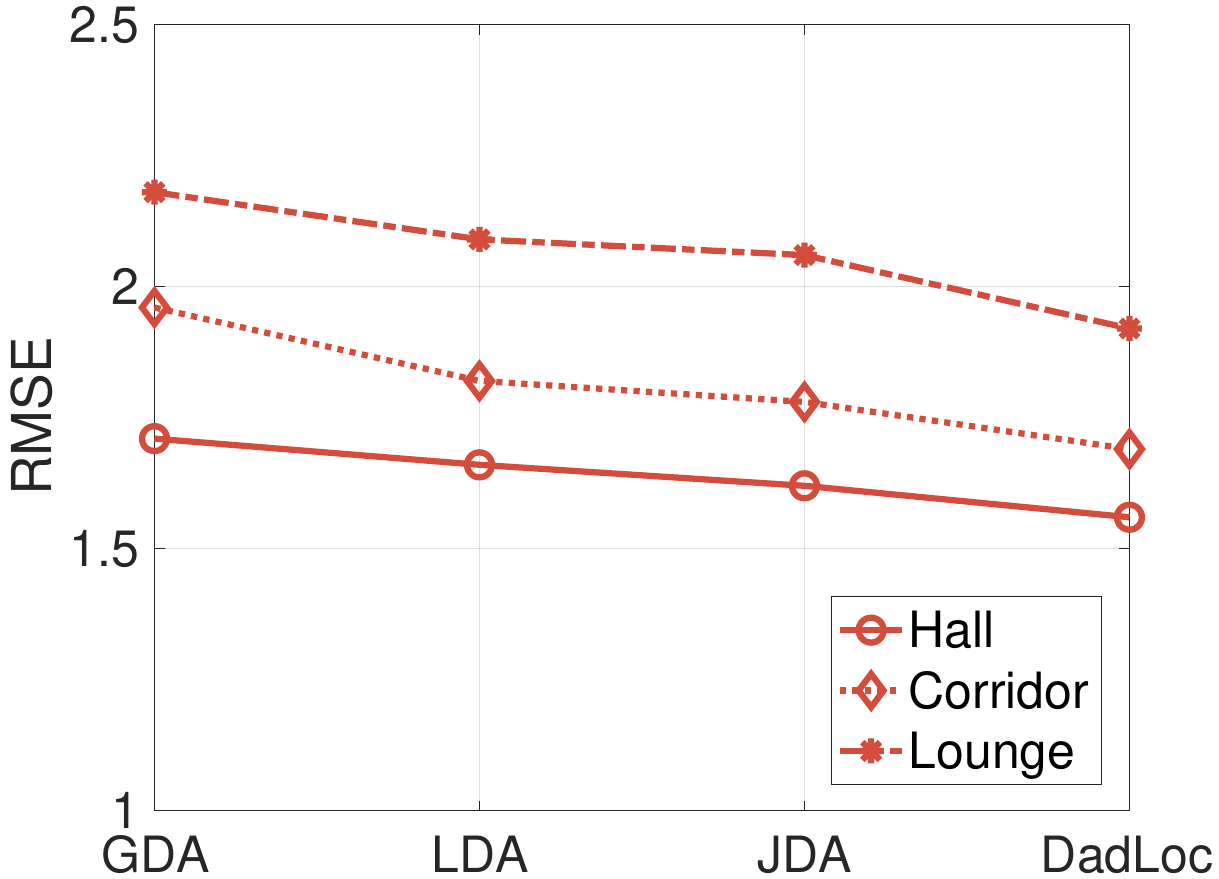}
	\caption{Performance Comparison on Different Distribution Adaptation }
	\label{fig_dda}
\end{figure}

To evaluate the effectiveness of dynamic distribution adaptation, we analyze the localization performance of the proposed DadLoc system with only global distribution adaptation (GDA), only  local distribution adaptation (LDA), and joint distribution adaptation (JDA) without the dynamics-adaptive factor, respectively. 
Fig. \ref{fig_dda} illustrates the localization errors of the proposed DadLoc  with different distribution adaptation schemes.
%The global distribution adaptation achieves a mean  error of $1.77m$ is relatively coarse than other schemes. 
The global distribution adaptation ($\mu=0$) achieves a mean  error of $1.95m$ by the confusion of the feature distributions across  different environments  without taking the location label information which  to some extent weakens the location discriminability of the feature representation. 
For the local distribution adaptation ($\mu=1$), the average error is reduced to $1.86m$ for robust localization.  The localization error of JDA with equal global and local domain adaptation ($\mu=0.5$) is the mean error of $1.82m$. The proposed DadLoc has a localization accuracy of $1.72m$ with dynamic distribution adaptation which attains an accuracy improvement of 11.8\% for GDA, 7.5\% for LDA, and 5.6\% for JDA, respectively. 
 These experimental results 
 indicate that it is necessary for robust localization to perform dynamic adversarial adaptation with complicated indoor environmental changes. %With the dynamic distribution adaptation,  
%the proposed DadLoc can achieve the average accuracy improvement of \% for GDA, \% for LDA, and \% for JDA, respectively. 
% With these localization results, 
The proposed DadLoc system is validated with the effectiveness of the developed DAAN by dynamic adversarial learning to learn finer environment-invariant  representations  for robust fingerprint-based localization with the trade-off between environmental adaptability and location discriminability.

\subsubsection{Effectiveness on Prediction Uncertainty Suppression}
\begin{figure}[thbp]
	\centering
 \includegraphics[width=0.4\textwidth]{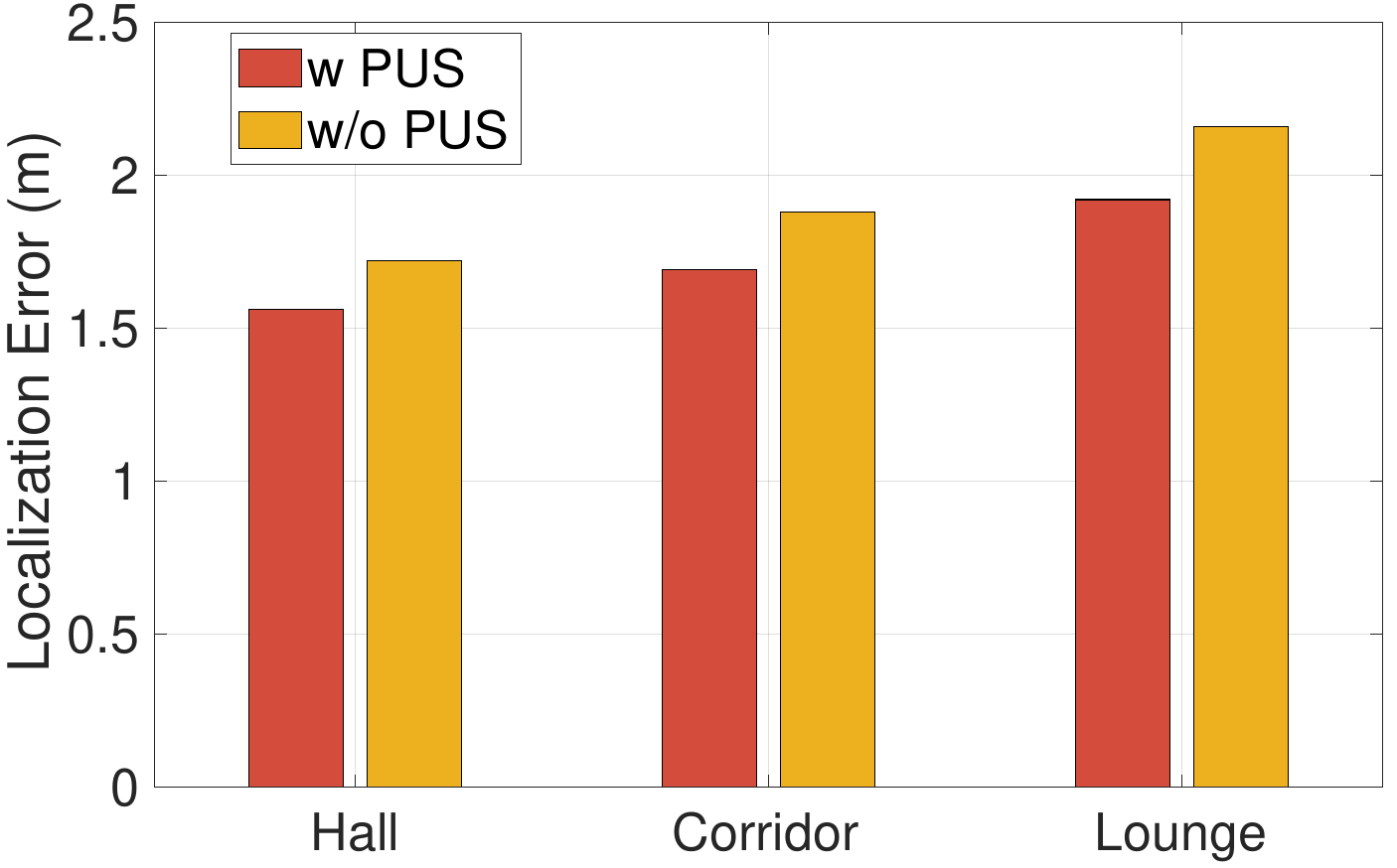}
	\caption{Localization Accuracy of  Prediction Uncertainty Suppression}
	\label{fig_pus}
\end{figure}

We further evaluate the effectiveness of  prediction uncertainty suppression (PUS) to guarantee effective and safe transfer with the developed DAAN. As shown in Fig. \ref{fig_pus}, the average accuracy of the proposed DadLoc shows superiority over the scheme without the strategy of prediction certainty suppression  by 10.6\% of the accuracy enhancement for  robust localization.   
Therefore, these experimental results show that it is effective for the proposed DadLoc to integrate the strategy of prediction uncertainty suppression into the developed DAAN, facilitating its localization performance for practical LBS. 
%conducted to reduce the localization error and effectively facilitate the DadLoc system performance of robust localization. 

\subsubsection{Parameter Analysis of the Developed DAAN}

\begin{figure*}[thbp]
	\centering
	\subfigure[$\lambda$]{
		\includegraphics[width=0.32\textwidth]{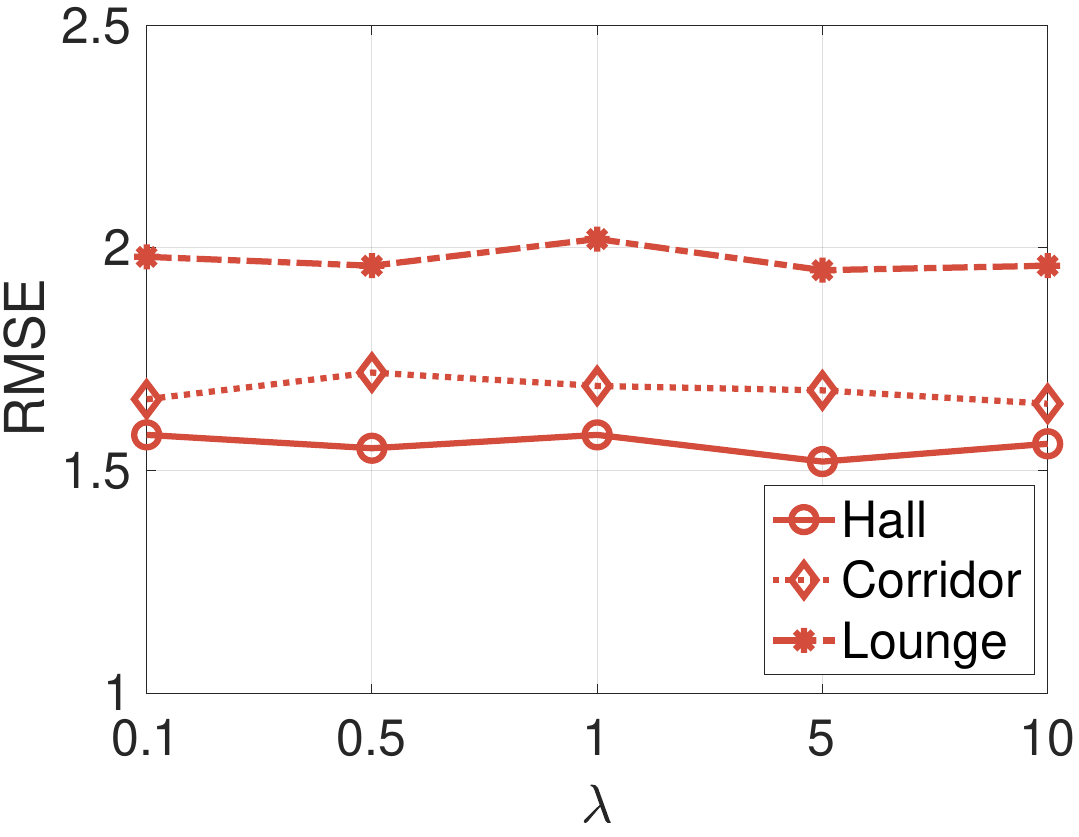}} 
		\label{lambda}
  \subfigure[$\gamma$]{
		\includegraphics[width=0.32\textwidth]{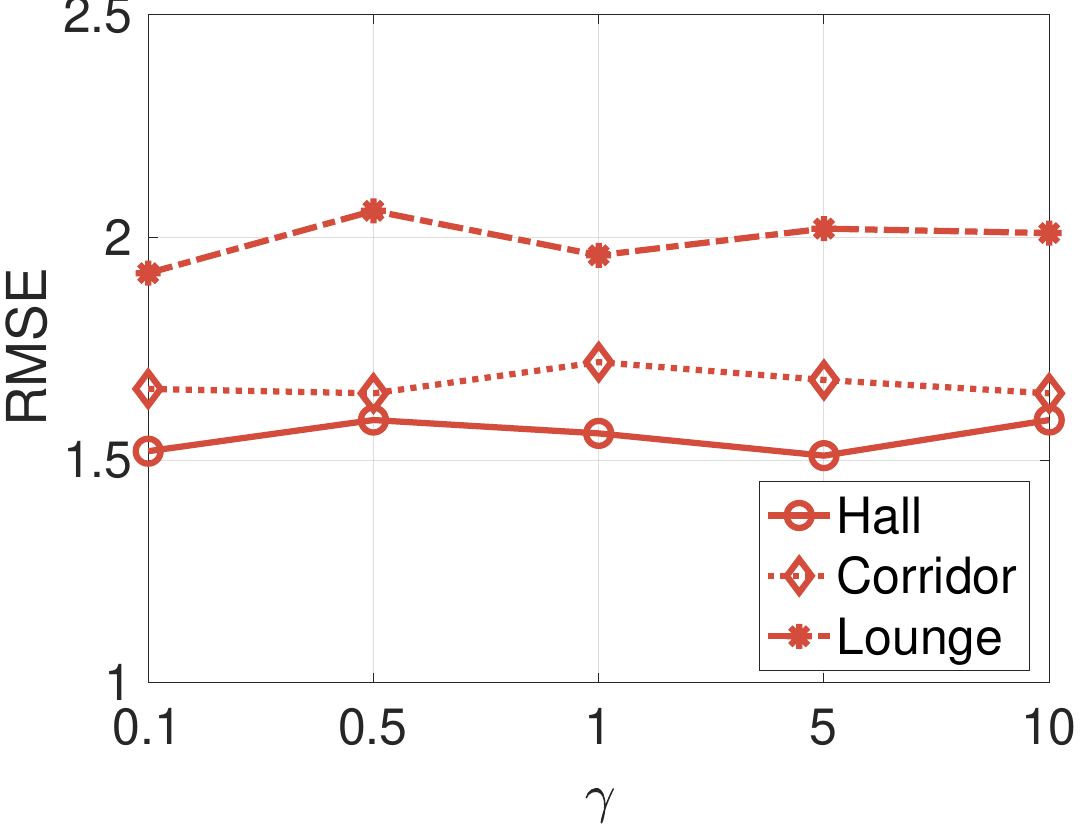}} 
		\label{gamma}
		\subfigure[Batch Size]{
			\includegraphics[width=0.32\textwidth]{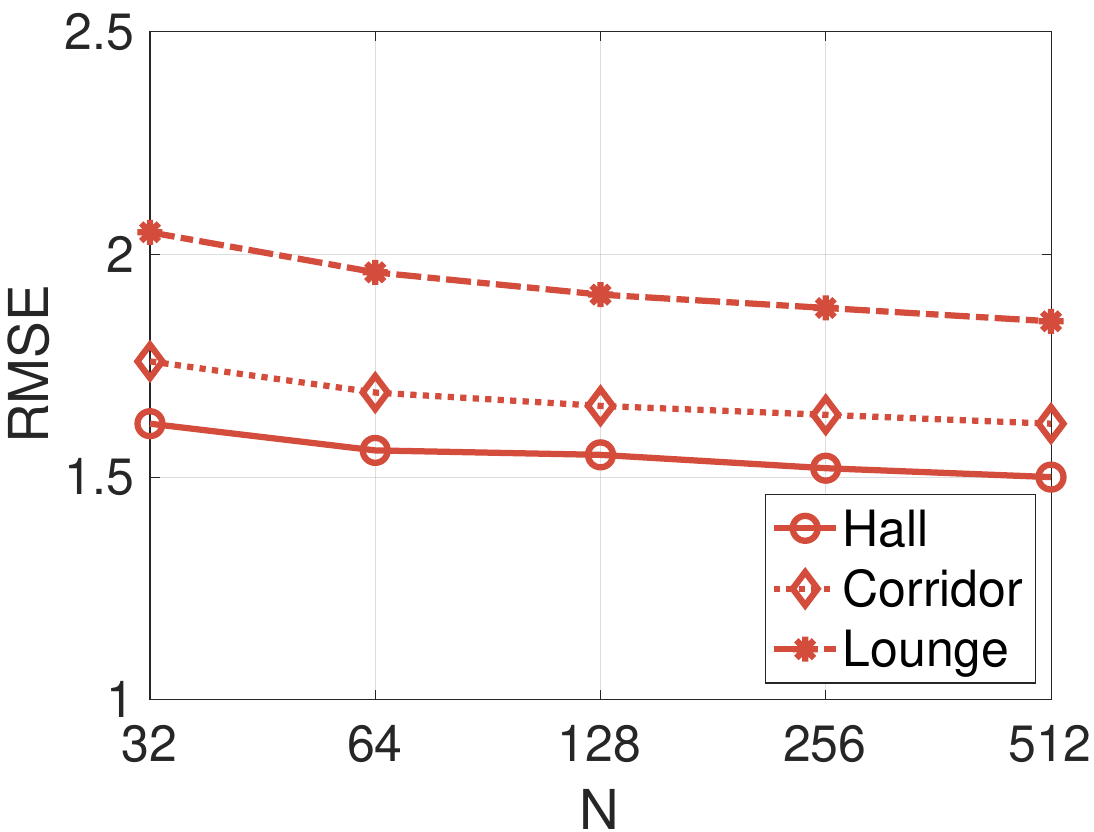}} 
		\label{batch}
		% \subfigure[Iteration]{
		% 	\includegraphics[width=0.22\textwidth]{iteration.pdf}} 
		% \label{iteration}
		\caption{Parameter Analysis of the Developed DANN}
	\label{fig_para}
\end{figure*}

The parameter analysis of the developed DAAN involves four meta-parameters, such as the penalty parameters $\gamma$ and $\lambda$, the batch size $N$, and the iteration number $T$ to control the training convergence of the developed DAAN. In Fig. \ref{fig_para}(a) and (b), the robust performance of the proposed DadLoc is realized with a wide range of parameter values for the penalty parameters, such as $\gamma$ and $\lambda$, respectively. As the illustration of Fig. \ref{fig_para}(c), the increasing amount of the batch size can improve the accuracy, but it is  unnecessary to make the batch size too large. The default value is set as 64 in our experiments. 
%Finally, we evaluate the convergence of the developed DAAN. 

\section{Conclusion}
In this paper, we propose a novel approach of robust fingerprint-based localization with dynamic adversarial learning, as DadLoc which incorporates multiple robust factors underlying RF fingerprints to learn  transferable representation with the developed DAAN. 
DadLoc conducts finer-grained distribution adaptation and quantifies the contributions of both global and local distribution adaptation in a dynamics-adaptive manner. We further employ the training strategy of prediction uncertainty suppression to attain the tradeoff between environmental adaptability and location discriminability of the learned deep representation for safe and effective feature transfer across different environments. 
 With extensive experimental results,  DadLoc achieves remarkable location performance improvement with an average localization error of $1.75m$ over other advanced works and classical deep-learning-based schemes.

% or


% \appendices


% \section{}

% \section*{Acknowledgment}

\ifCLASSOPTIONcaptionsoff
  \newpage
\fi

\end{document}